\documentclass[a4paper,11pt]{article}

\usepackage[margin=3cm]{geometry}

\usepackage[english]{babel}
\usepackage{amssymb}
\usepackage{amsmath}
\usepackage{amsthm}
\usepackage{psfrag}
\usepackage[T1]{fontenc}
\usepackage{ae,aecompl}
\usepackage[colorlinks]{hyperref}
\usepackage{subfigure}
\usepackage{appendix}
\usepackage{natbib}
\hypersetup{colorlinks,breaklinks=true,citecolor=blue,linkcolor=blue}
\usepackage{graphicx}
\usepackage{color}
\usepackage{url}

\providecommand\bnabla{\boldsymbol{\nabla}}

\newcommand\eg{e.g.\ }
\newcommand\ie{i.e.\ }

\newcommand{\pdt}[1]{\ensuremath{\frac{\mbox{$\partial$} #1}{\mbox{$\partial$} t}}}
\newcommand{\pl}{\left(}
\newcommand{\pr}{\right)}
\newcommand{\n}{\nabla}
\newcommand{\vn}{\boldsymbol{\nabla}}
\newcommand{\vect}[1]{\boldsymbol{#1}}
\newcommand{\vel}{\mathbf{u}}

\newcommand{\vorz}{\zeta}

\newcommand{\moyp}[1]{\overline{#1}}
\newcommand{\moyz}[1]{\left \langle #1 \right \rangle}

\newcommand{\zw}{\overline{u_{\phi}}}
\newcommand{\urms}{\mbox{\textit{Re}}}
\newcommand{\uzrms}{\mbox{\textit{Re}}_0}
\newcommand{\ucrms}{\mbox{\textit{Re}}_c}
\newcommand{\uzon}{U_0}
\newcommand{\us}{U_s}
\newcommand{\trms}{\Theta_{0}^{\ast}}

\newcommand{\Pran}{\mbox{\textit{Pr}}} 
\newcommand{\Ek}{\mbox{\textit{Ek}}}
\newcommand{\Ra}{\mbox{\textit{Ra}}}

\newcommand{\Ro}{\mbox{\textit{Ro}}}
\newcommand{\Nu}{\mbox{\textit{Nu}}}

\title{Subcritical convection of liquid metals in a rotating sphere using a quasi-geostrophic model}

\author{C\'eline Guervilly$^1$ and  Philippe Cardin$^{2}$ \vspace{0.2cm} 
\\ {\small $^1$School of Mathematics and Statistics, Newcastle University, Newcastle upon Tyne, NE1 7RU UK}
\\ {\small $^2$Institut des Sciences de la Terre, Universit\'e Grenoble Alpes, CNRS, 38041 Grenoble, France}}

\begin{document}

\maketitle

\begin{abstract}
We study nonlinear convection in a rapidly rotating sphere with internal heating  for values of the Prandtl number relevant for liquid metals
($\Pran\in[10^{-2},10^{-1}]$).
We use a numerical model based on the quasi-geostrophic approximation, in which variations of the axial vorticity along the rotation axis are neglected, whereas
the temperature field is fully three-dimensional. 
We identify two separate branches of convection close to onset: (i) a well-known weak branch for Ekman numbers greater than $10^{-6}$, 
which is continuous at the onset (supercritical
bifurcation) and
consists of thermal Rossby waves, and (ii) a novel strong branch at lower Ekman numbers, which is discontinuous at the onset. 
The strong branch becomes subcritical for Ekman numbers of the order of $10^{-8}$.
On the strong branch, the Reynolds number of the flow is greater than $10^3$,
and a strong zonal flow with multiple jets develops, even close to the nonlinear onset of convection.
We find that the subcriticality is amplified by decreasing the Prandtl number.
The two branches can co-exist for intermediate Ekman numbers, leading to hysteresis ($\Ek=10^{-6}$, $\Pran=10^{-2}$).
Non-linear oscillations are observed near the onset of convection for $\Ek=10^{-7}$ and $\Pran=10^{-1}$.
\end{abstract} 

\section{Introduction}
\label{sec:intro}
Thermal convection of a Boussinesq fluid in a rotating sphere is a classical problem in 
fluid dynamics that has important implications for the heat transfer and generation of magnetic fields in astrophysical bodies.
In the liquid core of planets, the Ekman number, $\Ek$, which measures the viscous
effects compared with the Coriolis force, and the Rossby number, $\Ro$, which measures the nonlinear inertial
effects compared with the Coriolis force, are very small, so convection is
strongly influenced by the global rotation.  
The linear onset of thermal convection in spherical geometry for $\Ek\ll1$ and $\Ek/\Pran\ll1$ (where
$\Pran$ is the Prandtl number, the ratio of the kinematic viscosity to thermal diffusivity of the fluid) 
is now well understood following the theoretical
work of \citet{Rob68,Bus70,Sow77,Yan92,Jon00}, confirmed by experimental studies \citep{Car83, Cor92, Car94} 
and numerical models \citep{Or87, Zha92, Zha93b, Ard97, Dor04}.
At the linear onset, the thermal instability takes the form of 
columnar flows aligned with the rotation axis, 
whose axial vorticity has much smaller variations along the rotation axis than in the azimuthal and 
cylindrical radial directions
to accommodate the Proudman-Taylor theorem.
These convective columns drift in the azimuthal direction, analogously to Rossby waves, and are therefore called thermal Rossby waves.

The properties of nonlinear convection for small $\Ek$ and $\Ro$ have received considerable attention 
but are still not fully understood (\eg \citet{Jon07,Aur15} for a review).
At the onset of instability, any infinitesimal perturbation grows exponentially until nonlinear effects saturate the growing 
solution. If the transition from the basic conduction state to the convection state is continuous, 
then we have a supercritical branch of convection \citep[\eg][]{Man04}. 
However, in some classical problems of fluid dynamics, the nonlinearities can promote the instability rather than saturating it.
This leads to a subcritical bifurcation, meaning that, once the instability develops above the linear onset, 
it might continue to exist when the controlling parameter is decreased below the critical value for linear instability. 
Subcritical behaviour has been observed in various convective systems such as 
double-diffusive convection \citep[\eg][]{Dac81}, planar convection with moderate rotation \citep{Ver66,Baj02},
and B\'enard-Marangoni convection \citep[\eg][]{Sca67}.  
In thermal convection in rotating spherical geometry, weakly nonlinear analyses
show that the nature of the bifurcation from the basic conduction state to the thermal Rossby waves 
might change from supercritical to subcritical in the limit of small Ekman numbers, at least in the case of internal heating \citep{Sow77,Pla08}.
However, nonlinear numerical models, which can be computed down to $\Ek=\textit{O}(10^{-6})$, 
always find that the bifurcation is supercritical \citep[\eg][]{Zha92, Aub03, Gro01, Chr02, Mor04, Yad16}.
This possibly means that the numerical models are not run at sufficiently small Ekman numbers to properly capture the properties of nonlinear convection
for planetary cores (where $\Ek=\textit{O}(10^{-15})$), even near the onset of convection. 
Current numerical models might thus misrepresent realistic planetary convection 
for more vigorous thermal driving as well. The scaling laws for the global properties of convection
that are derived from the numerical models \citep[\eg][]{Chr02} would then not be applicable for natural objects.
Furthermore, calculations are generally performed for Prandtl numbers  close to unity 
while liquid metals have smaller $\Pran$; for instance, the liquid iron in the Earth's core is thought to have $\Pran\approx10^{-1}$ \citep{Poz12}.
Subcriticality is observed for small Prandtl numbers in planar convection with moderate rotation \citep{Ver66,Cle00}, and the value of $\Pran$ might 
be equally significant for the nature of the bifurcation at onset in rapidly-rotating spherical convection.

The objective of the present paper is to describe nonlinear convection for $\Ek\ll1$, $\Ek/\Pran\ll1$ and $\Ro\ll1$ near the onset. 
In order to map the different hydrodynamical regimes, we carry out an extensive exploration of the parameter space.  
To alleviate part of the computational limitations that apply to global models of planetary cores, 
we use the quasi-geostrophic (QG) approximation that was developed 
by \citet{Bus86} for thermal convection in the annulus geometry of \citet{Bus70} with curved boundaries, 
and subsequently used and modified for the spherical geometry
by \citet{Car94,Aub03,Mor04,Sch05,Gil06,Gil07,Pla08,Cal12}.
The model assumes that the flow is geostrophically balanced, 
\ie the Coriolis force balances the pressure gradient at leading order.
By neglecting the variation of the axial vorticity along the rotation axis, 
we can model the flow in two dimensions. 
This is an important limitation to the full dynamics of rotating convection, but this exploratory study aims to inform future three-dimensional studies.
The QG approximation allows us to model thermal convection for Ekman numbers as low as $10^{-8}$.
Results from the numerical implementation of the QG approximation in spherical geometry have been successfully
benchmarked against the asymptotic theory \citep{Gil06,Lab15} and 3D numerical models \citep{Aub03,Pla08} for the onset of convection, and against 
laboratory experiments in rotating convection \citep{Aub03, Gil07} and in shear flows \citep{Sch05} for the nonlinear regime.

The main issue of the QG approximation for rotating spherical convection is that the basic temperature background
has a spherical symmetry. Treating the temperature as a 2D field is therefore an unjustified approximation, although it has been successfully used  
in previous studies  when compared with results from 3D models and laboratory experiments \citep{Aub03, Mor04, Gil07, Pla08}.
In the present study, we follow the work of \citet{Sch06} by coupling the 2D velocity to a 3D implementation of the temperature in the whole sphere. 
The coupling terms require interpolations from the 2D and 3D grids.
The  hybrid QG-3D approach can therefore only be efficient, in term of the computational time, if
the temperature is solved on a 3D grid that is coarser than the 2D grid used for the velocity.
This is particularly appropriate to model fluids with small Prandtl numbers, where the dissipative scale of the temperature is larger than
the viscous dissipation scale. 
Recent studies of rapidly-rotating convection in small Prandtl number fluids have shown that the Prandtl number has a significant influence on the convection,
especially on the dominant length and time scales of the flow \citep{Gil07,Cal12,Kin13}.  
In this paper,  we vary $\Pran$ between $[10^{-2},10^{-1}]$. 
Note that for small Prandtl numbers, convection cells attached to the equator of the 
outer sphere are preferred to thermal Rossby waves at the linear onset of convection 
when the Ekman number is moderately small \citep{Zha87,Bus04,Pla05,San16}. 
In the limit $\Ek\ll1$ and $\Ek/\Pran\ll1$, the critical Rayleigh number at the onset of these equatorially-attached modes scales as $\Ek^{-2}$, 
while the critical Rayleigh number at the onset of the thermal Rossby waves scales as $\Ek^{-4/3}$ \citep{Bus04}. The equatorially-attached 
modes are therefore unlikely to occur in the Earth's core. The 3D numerical simulations of \citet{Ard97} show that the thermal Rossby waves are 
preferred for $\Ek\leq10^{-4}$ when $\Pran=10^{-1}$ and $\Ek\leq10^{-6}$ when 
$\Pran=10^{-2}$. 
The equatorially-attached modes are outside the scope of this paper, so we restrict our study to 
this range of small Ekman numbers.

For simplicity, we consider only thermal convection in a full sphere without a solid inner core to avoid the singularity of the QG approach 
at the tangent cylinder. Consequently, the thermal convection is driven by an homogenous internal heating, 
which is a well-studied variation of the classical problem of thermal convection in a rapidly-rotating sphere \citep{Jon00},
and is relevant for the early history of the Earth's core \citep[\eg][]{Ols13}.

The layout of the paper is as follows.
First, we derive the formulation of the hybrid QG-3D model that describes thermal convection 
in a rapidly rotating sphere in \S\ref{sec:model}.
Then in \S\ref{sec:NL}, we discuss the results from the nonlinear simulations carried out near the onset
for small Ekman numbers. 
Subcritical convection, hysteresis and nonlinear oscillations are reported.  
A discussion of the results is given in \S\ref{sec:ccl}.
In Appendix~\ref{sec:linear}, the results at the linear onset of convection obtained with our hybrid model are
compared with results from the asymptotic theory and from previous quasi-geostrophic and 3D numerical models.

\section{Formulation of the model}
\label{sec:model}
We study thermal Boussinesq convection in a rotating sphere driven by internal heating.
The system rotates at a constant angular velocity \mbox{$\Omega \vect{e}_z$}.
The acceleration due to gravity is radial and linear, \mbox{$\vect{g}=g_0 r \vect{e}_r$}.
The radius of the sphere is $r_o$ and no inner core is present.
The fluid has kinematic viscosity $\nu$, thermal diffusivity $\kappa$, density $\rho$, heat capacity 
at constant pressure $C_p$, and thermal expansion coefficient $\alpha$, all of which are constant.
We consider an homogeneous internal volumetric heating $S$. 
In the absence of convection, the static temperature profile $T_s$ is calculated by solving the diffusive heat 
equation and can be written as
\begin{equation}
	T_s (r) = T_o + \frac{S}{6\kappa \rho C_p} (r_o^2-r^2),
	\label{eq:Ts}
\end{equation}
where $T_o$ is the imposed temperature at the boundary, $r=r_o$. 
The governing equations are solved in dimensionless form, obtained by scaling lengths
with $r_o$, times with $r_o^2/\nu$, and temperature with \mbox{$\nu S r_o^2/(6\rho C_p\kappa^2)$}.
The system of dimensionless equations is:  
\begin{eqnarray}
	&&\pdt{\vel} + \pl \vel \cdot \boldsymbol{\n} \pr \vel + \frac{2}{\Ek}\mathbf{e}_z \times \vel 
	= - \boldsymbol{\n} p + \boldsymbol{\n^2} \vel + \Ra \Theta \mathbf{r} ,
	\label{eq:NS1}
	\\
	&& \nabla \cdot \vel = 0,
	\\
	&& \pdt{\Theta} + \vel \cdot \nabla \Theta -\frac{2}{\Pran} r u_r = \frac{1}{\Pran} \nabla^2 \Theta,
	\label{eq:T1}
\end{eqnarray}
where $\vel$ is the velocity field, $p$ the modified pressure, which includes the centrifugal potential, and
$\Theta$ the temperature perturbation relative to the static temperature~(\ref{eq:Ts}). 

The dimensionless numbers are, the Ekman number,
\begin{equation}
	\Ek=\frac{\nu}{\Omega r_o^2}, 
\end{equation}
the Rayleigh number,
\begin{equation}
	\Ra=\frac{\alpha  g_0 S r_o^6}{6 \rho C_p \nu \kappa^2},
\end{equation}
and the Prandtl number,
\begin{equation}
	\Pran=\frac{\nu}{\kappa}.
\end{equation}

At $r=r_o$, the boundary condition for the velocity is no-slip and impenetrable and 
the temperature is fixed, 
\begin{equation}
	\vel = \vect{0}, \quad  \Theta = 0 \textrm{ at } r=r_o.
\end{equation}

We detail below the quasi-geostrophic formulation used to model
the velocity field, the 3D model used for temperature, and the numerical method.
Throughout this paper, we use both spherical coordinates $\vect{u}=(u_r,u_{\theta},u_{\phi})$ and 
cylindrical polar coordinates $\vect{u}=(u_s,u_{\phi},u_{z})$.

\subsection{Governing equation for the non-axisymmetric flow}
We seek to model the system of equations~(\ref{eq:NS1})-(\ref{eq:T1}) for small Ekman numbers. 
We choose the quasi-geostrophic (QG) approximation to model the evolution of the velocity field 
\citep{Or87,Bru93,Car94,Pla02,Aub03,Jon03,Mor04,Sch05,Gil06,Cal12,Tee12}.
The QG approximation reduces the three-dimensional system to a two-dimensional system 
by taking advantage of the small variation of the flow along $z$ due to the rapid rotation \citep{Gil06}.
This approximation is only justified in the case of small slope of the boundaries, such as
the \citet{Bus70} annulus. In the case of a sphere, the approximation is therefore
not rigorously justified in any asymptotic limit. 
Consequently, the QG model is intended as a simplified model of convection in a rapidly rotating sphere
that allows us to investigate unexplored regions of the parameter space.
When possible, comparisons with theoretical, experimental and 3D numerical models
show that the QG model correctly reproduces key properties of the full system \citep{Aub03,Mor04,Gil06,Gil07,Pla08}.

The QG model assumes that the fluid dynamics is dominated by the geostrophic balance, \ie the
Coriolis force balances the pressure gradient. The geostrophic velocity $\vel^g$
is invariant along $z$ and $\vel^g = (u^g_s,u^g_{\phi},0)$ in cylindrical polar coordinates.
By taking the $z$-component of the curl of the momentum equation~(\ref{eq:NS1}) and averaging it
along $z$, we obtain the equation for the axial vorticity, $\vorz^g = \pl \bnabla \times \vel^g \pr \cdot \vect{e}_z$,
\begin{equation}
	\pdt{\vorz^g} + \pl \vel^g \cdot \bnabla \pr \vorz^g 
	- \pl \frac{2}{\Ek} + \vorz^g \pr \moyz{\frac{\partial u_z}{\partial z}}
	= \nabla_e^2 \vorz^g - \Ra \moyz{\frac{\partial \Theta}{\partial \phi}},
	\label{eq:vorz}
\end{equation}

with 

\begin{equation}
	\nabla^2_e A  \equiv \frac{1}{s} \frac{\partial}{\partial s}\pl s \frac{\partial A}{\partial s} \pr
					+ \frac{1}{s^2} \frac{\partial^2 A}{\partial \phi^2},
\end{equation}
and 
\begin{equation}
 \moyz{A}  \equiv \frac{1}{2H} \int^{+H}_{-H} A dz,
\end{equation}
where $H=\sqrt{1-s^2}$ is the axial distance from the spherical boundary to the equatorial plane.

The axial velocity $u_z$ is assumed to be linear in $z$. 
Three-dimensional linear convection models with internal 
heating show that the variation in $z$ of $u_z$ is not exactly linear, 
but it is a reasonable assumption because the variations in $z$ with respect to the linear profile are small 
compared with the variations in $s$ and $\phi$ \citep{Gil06}. 

The velocity $\vel^g$ can be described by a streamfunction $\psi$ that describes the 
non-axisymmetric (\ie $\phi$-dependent) geostrophic flow with the addition of an axisymmetric azimuthal flow,
\begin{equation}
	\vel^g = \frac{1}{H} \bnabla \times \pl H \psi \vect{e}_z \pr + \moyp{u_{\phi}^g} \vect{e}_{\phi},
\end{equation}
where
\begin{equation}
 \moyp{A}  \equiv \frac{1}{2\pi} \int_{0}^{2\pi}A d\phi.
\end{equation}
This choice for the streamfunction implies that the divergence of $\vel^g$ in the equatorial plane is non-zero because of
the return axial flow due to the slope of the boundaries, 
\begin{equation}
	\bnabla_e \cdot \vel^g = -\beta u_s^g,
\end{equation}
where
\begin{equation}
	\bnabla_e \cdot  \vect{A} \equiv \frac{1}{s} \frac{\partial s A_s}{\partial s} + \frac{1}{s} \frac{\partial A_{\phi}}{\partial \phi},
\end{equation}
and
\begin{equation}
	\beta = \frac{1}{H}\frac{dH}{ds} = - \frac{s}{H^2}.
\end{equation}

The third term on the left-hand side of equation~(\ref{eq:vorz}) requires us to determine 
 $u_z$ at the boundary $z=\pm H$: 
\begin{equation}
	\left. u_z\right|_{\pm H}   = \pm \frac{1}{H} \left. \vel \cdot \mathbf{n}\right|_{\pm H} 
	\pm \beta H u_s^g ,
	\label{eq:uzH}
\end{equation}
where the normal vector at the boundary is $\vect{n} = \vect{e}_r$.
The normal component, $\left. \vel \cdot \mathbf{n}\right|_{\pm H}$, is the Ekman pumping
induced by the viscous boundary layer and is determined by asymptotic methods \citep{Gre68},
\begin{equation}
  \left. \vel \cdot \mathbf{n}\right|_{z=\pm H} = -\frac{\Ek^{1/2}}{2} \mathbf{n} \cdot \vn \times \pl \frac{\mathbf{n}\times \vel \pm \vel}{\sqrt{\left| \mathbf{n}\cdot \mathbf{e_z} \right|}} \pr 
   = \Ek^{1/2} P(s,u_s^g, u_{\phi}^g).
\end{equation}
This analytical formulation is valid for a linear Ekman layer and for variations of the velocity that 
are slower than the rotation period and on larger lengthscales than $\mathit{O}(\Ek^{1/2})$.
The function $P$ is derived for a spherical boundary in \citet{Sch05}.

Our numerical code solves the equation for the evolution of the non-axisymmetric streamfunction  
\begin{equation}
	\pdt{}\mathcal{L}\psi + \pl \vel^g \cdot \bnabla_e \pr \mathcal{L}\psi 
	- \pl \frac{2}{\Ek} + \mathcal{L}\psi  \pr \pl \frac{\Ek^{1/2}P}{2H^2}  + \beta u_s^g \pr
	= \nabla_e^2 \mathcal{L}\psi  - \Ra \moyz{\frac{\partial \Theta}{\partial \phi}},
	\label{eq:psi}
\end{equation}
where
\begin{equation}
	\mathcal{L}\psi  = \vorz^g = 
	-\nabla_e^2 \psi + \frac{1}{s}\frac{\partial}{\partial s} \pl \frac{s^2\psi}{H^2} \pr.
\end{equation}

The no-slip and impenetrable boundary conditions imply that
\begin{equation}
	\psi = \frac{\partial \psi}{\partial s} = 0 \textrm{ at } s=1.
	\label{bc:psi1}
\end{equation}
At the centre of the sphere, the solution must be regular.
Using a decomposition in Fourier modes,
\begin{equation}
	\psi(s,\phi,t) = \sum_{m=1}^{+\infty} \hat{\psi}^m (s,t) e^{im\phi},
\end{equation}
the regularity condition is 
\begin{equation}
	\hat{\psi}^m =  \textit{O}(s^m) \textrm{ at } s=0.
	\label{bc:psi0}
\end{equation}

\subsection{Governing equation for the zonal flow}
In our model, the streamfunction $\psi$ only describes the non-axisymmetric motions, so the axisymmetric 
azimuthal flows, or zonal flows, are treated separately.
We take the $\phi$- and $z$-averages of the $\phi$-component of the momentum equation to obtain
\begin{equation}
 \pdt{\moyp{u_{\phi}^g}}
 +   \moyp{u_{s}^g \frac{\partial u_{\phi}^g}{\partial s}} + \moyp{\frac{u_{s}^g u_{\phi}^g}{s}}
 +\frac{2}{\Ek} \left \langle \moyp{u_s} \right \rangle
 = \n^2 \moyp{u_{\phi}^g} - \frac{\moyp{u_{\phi}^g}}{s^2} .
\label{eq:uzonal}
\end{equation}

Note that the geostrophic balance imposes that $\moyp{u_{s}^g}=0$.
The fourth term on the left-hand side of (\ref{eq:uzonal}) involves the $z$-dependent radial velocity, 
which corresponds to the Ekman pumping term. 
Using the incompressibility of the fluid, it can be shown \citep{Aub03} that
\begin{equation}
 \left \langle \moyp{u_s} \right \rangle= 
 	\frac{\Ek^{1/2}}{2H^{3/2}} \moyp{u_{\phi}^g} .
\end{equation}

The no-slip boundary condition at the outer sphere implies
\begin{equation}
	\moyp{u_{\phi}^g} = 0 \textrm{ at } s=1.
	\label{bc:up1}
\end{equation}
By symmetry at the centre,
\begin{equation}
	\moyp{u_{\phi}^g} = 0 \textrm{ at } s=0.
	\label{bc:up0}
\end{equation}

\subsection{Governing equation for the temperature}

The dimensionless equation for the evolution of the temperature perturbation in 3D is
\begin{equation}
 \pdt{\Theta}+ \vel^{3d} \cdot \boldsymbol{\n} \Theta = 
	\frac{1}{\Pran} \pl 2 r u^{3d}_r+\n^2 \Theta \pr.
	\label{eq:T}
\end{equation}
where $\vel^{3d}$ is the velocity in 3D. 
In cylindrical polar coordinates, 
\begin{equation}
	\vel^{3d} = (u_s^g, u_{\phi}^g,\Ek^{1/2} z P + \beta z u_s^g) .
	\label{eq:u3d}
\end{equation}

The temperature is fixed at the outer boundary so 
\begin{equation}
	\Theta = 0 \textrm{ at } r=1.
	\label{bc:T1}
\end{equation}
At the centre, the heat flux of the spherically averaged temperature perturbation is zero,
\begin{eqnarray}
	\left \langle \frac{\partial \Theta}{\partial r} \right \rangle_\mathcal{S} =
	  \int_{\mathcal{S}} \frac{\partial \Theta}{\partial r} d\mathcal{S} = 0 \textrm{ at } r=0,
	\label{bc:T0a}
\end{eqnarray}
where $\mathcal{S}$ is a spherical surface.
By symmetry, the non-spherically symmetric components of $\Theta$ are zero,
\begin{equation}
	\Theta -  \left \langle \Theta \right \rangle_\mathcal{S} = 0 \textrm{ at } r=0.
	\label{bc:T0b}
\end{equation}

\subsection{Numerical method}
In summary, the governing equations of our hybrid quasi-geostrophic-3D model are
\begin{eqnarray}
	&& \pdt{}\mathcal{L}\psi + \pl \vel \cdot \bnabla_e \pr \mathcal{L}\psi 
	- \pl \frac{2}{\Ek} + \mathcal{L}\psi  \pr \pl \frac{\Ek^{1/2}P}{2H^2}  + \beta u_s \pr
	= \nabla_e^2 \mathcal{L}\psi  - \Ra \moyz{\frac{\partial \Theta}{\partial \phi}},
	\label{eq:psi_f}
	\\
 	&& \pdt{\moyp{u_{\phi}}}
 	+  \moyp{u_{s} \frac{\partial u_{\phi}}{\partial s}} + \moyp{\frac{u_{s} u_{\phi}}{s}}
	 + \frac{\moyp{u_{\phi}}}{\Ek^{1/2} H^{3/2}} 
	 = \n^2 \moyp{u_{\phi}} - \frac{\moyp{u_{\phi}}}{s^2} ,
	\label{eq:uzonal_f}
	\\
	 &&\pdt{\Theta}+ \vel^{3d} \cdot \boldsymbol{\n} \Theta = 
	\frac{1}{\Pran} \pl 2 r u^{3d}_r+\n^2 \Theta \pr,
	\label{eq:T_f}
\end{eqnarray}
subject to the boundary conditions~(\ref{bc:psi1}), (\ref{bc:psi0}), 
(\ref{bc:up1})-(\ref{bc:up0}), (\ref{bc:T1})-(\ref{bc:T0b}), and the 3D velocity used in the temperature equation is given by 
equation~(\ref{eq:u3d}).
The superscripts $g$ have been removed for clarity. 

Equations~(\ref{eq:psi_f}) and (\ref{eq:uzonal_f}) are solved on a two dimensional grid 
in the equatorial plane. A 
second-order finite difference scheme is implemented in radius with irregular spacing (finer
near the outer boundary). In the azimuthal direction, the streamfunction
and zonal velocity are expanded in Fourier modes. 
Equation~(\ref{eq:T_f}) is solved on a three dimensional grid. 
Similarly to the 2D grid, a finite difference scheme is used in radius. 
The temperature is expanded in spherical harmonics $Y_l^m$ in the angular coordinates with $l$
representing the latitudinal degree and $m$ the azimuthal mode. 
This allows for a simple implementation of the boundary conditions at the centre.
 A Crank-Nicolson scheme is implemented for the time integration of the diffusion terms and an 
 Adams-Bashforth procedure is used for the other terms.
 Our hybrid approach is numerically efficient compared with a fully 3D approach
 when considering small Prandtl numbers, \ie when the diffusion of the temperature happens on shorter timescales than 
 the viscous dissipation. In this case, the 3D grid of the temperature requires less radial points 
and azimuthal modes than the 2D grid of the velocity, so smaller gridsteps and timestep can be used for the velocity.
The most demanding nonlinear simulations were run at $\Ek=10^{-8}$.
The numerical resolution required to resolve the simulation run at $\Ek=10^{-8}$ and $\Pran=10^{-1}$ for the largest
Rayleigh number ($\Ra\approx 6\Ra_c$)
is $1600$ radial points, $300$ Fourier modes $m$
for the 2D velocity grid and $800$ radial points, $200$ degrees $l$ and $200$ modes $m$
of the spherical harmonics for the 3D temperature grid.

The buoyancy term in equation~(\ref{eq:psi_f}) requires the integration of the temperature along $z$
for each point of coordinates $(s,\phi)$
on the 2D grid at each timestep. To do so, we use a linear interpolation in $r$ and $\theta$
from the 3D spherical grid onto a 3D cylindrical projection of the velocity grid. The 
spectral decomposition in $\phi$ for the velocity and temperature 
allows a straightforward interpolation in spectral space.
To calculate the 3D velocity used for the advection of temperature, we 
use a linear interpolation in $s$. Again, the interpolation in $\phi$ is straightforward
in spectral space.

In order to benchmark the numerical model,
we performed calculations with the linearised version of the code to determine the critical parameters at the linear onset of convection. 
The linear results are given in Appendix~\ref{sec:linear} and compared with results from the asymptotic 
theory and from published numerical studies using either a 3D or a QG model. The comparison shows that our 
numerical results tend to the asymptotic values as the Ekman number decreases.
In the following, the critical Rayleigh number at the linear onset of convection calculated with our 
numerical code is denoted $\Ra_c$.

\subsection{Definition of the output parameters}
To quantify the global properties of convection, we use the following quantities calculated from the output of the simulations.  
In our dimensionless units, the Reynolds number corresponds to the root mean square (rms) value of the velocity,
\begin{equation}
	\urms^{\ast}(t) = \pl \frac{3}{4\pi} \int_0^{2\pi} \int_0^{1} (u_s^2(s,\phi,t) + u_{\phi}^2(s,\phi,t)) 2 H(s) s ds d\phi \pr^{1/2},
	\label{eq:def_Re}
\end{equation}
and 
 \begin{equation}
	\urms = \frac{1}{\Delta t} \int_{\Delta t} \urms^{\ast}(t) dt,
 \end{equation}
where $u_{\phi}$ includes the zonal velocity. 
We measure the convective Reynolds number, $\ucrms^{\ast}$, as in equation~(\ref{eq:def_Re})
but including only the non-axisymmetric velocity. Its time average is denoted $\ucrms$.
Similarly, we measure the zonal Reynolds number, $\uzrms^{\ast}$, including only the axisymmetric velocity, and its time average is denoted $\uzrms$.
The time averages are calculated over a long period of time after the system reaches a statistical equilibrium. 

We define the radial profile of the time-averaged zonal velocity as
\begin{equation}
	\uzon(s) = \frac{1}{\Delta t}\int_{\Delta t} \moyp{u_{\phi}}(s,t) dt,
\end{equation}
and the radial profile of the rms radial velocity,
\begin{equation}
	\us(s) = \frac{1}{\Delta t} \int_{\Delta t} \pl \frac{1}{2\pi s} \int_0^{2\pi} u_{s}^2(s,\phi,t) s d\phi \pr^{1/2} dt.
\end{equation}

The Nusselt number, which is the ratio of the total heat flux, $Q_{t}$, to the conductive heat flux $Q_{\kappa}$, is 
often used to measure the efficiency of the convective heat transport. 
In a convective system with internal heating and fixed temperature boundaries,
the mean total outward heat flux must balance heat production, so $Q_{t}=r_oS/3$.
To estimate the heat flux that is conducted along the spherically averaged temperature gradient,
we define the equivalent internal heating, $S_{eq}$, that would correspond to the purely conductive 
state with a temperature at the centre 
$T_c=T_s(r=0)+\left \langle \Theta(r=0) \right \rangle_{\mathcal{S}}$, where 
$\left \langle \Theta(r=0) \right \rangle_{\mathcal{S}}$ is an output of the simulation. 
Assuming that $Q_{\kappa} = r_oS_{eq}/3$,
we can define the Nusselt number, $\Nu$, as
\begin{equation}
	 \Nu= \frac{Q_t}{Q_{\kappa}} = \frac{S}{S_{eq}} = \frac{1}{\Pran T_c},
	\label{eq:Nu}
\end{equation}
where $S_{eq} = \Pran T_c S $ and $T_c=\Pran^{-1} + \left \langle \Theta(r=0) \right \rangle_{\mathcal{S}}$.
The decrease of the temperature at the centre due to convection is thus a good measure of the efficiency of the
convective heat transport \citep{Gol16}.
We find that longer time integrations are required to ensure the saturation of $\Nu$ than 
for the saturation of the kinetic energy.

We also measure the rms value of the axisymmetric temperature perturbation as 
\begin{equation}
	\trms(t) = \pl \frac{3}{2} \int_0^1 \int_0^{\pi} \moyp{\Theta}^2 r^2 \sin\theta dr d\theta \pr^{1/2}.
\end{equation}

\section{Non-linear calculations near the onset of convection}
\label{sec:NL}

The data set presented in this paper contains approximately 120 simulations performed at 
small Ekman numbers, $\Ek\in[10^{-8},10^{-5}]$, and small Prandtl numbers, $\Pran\in[10^{-2},10^{-1}]$.
The focus of the study is convection near the onset, so 
the Rayleigh number has been varied up to 10 times above the critical value at the onset of convection.
The values of the critical Rayleigh number at the linear onset, $\Ra_c$, are given in table~\ref{tab:critpar}
in Appendix~\ref{sec:linear} for different $\Ek$ and $\Pran$.

\begin{figure}
\centering   
   \includegraphics[clip=true,width=14cm]{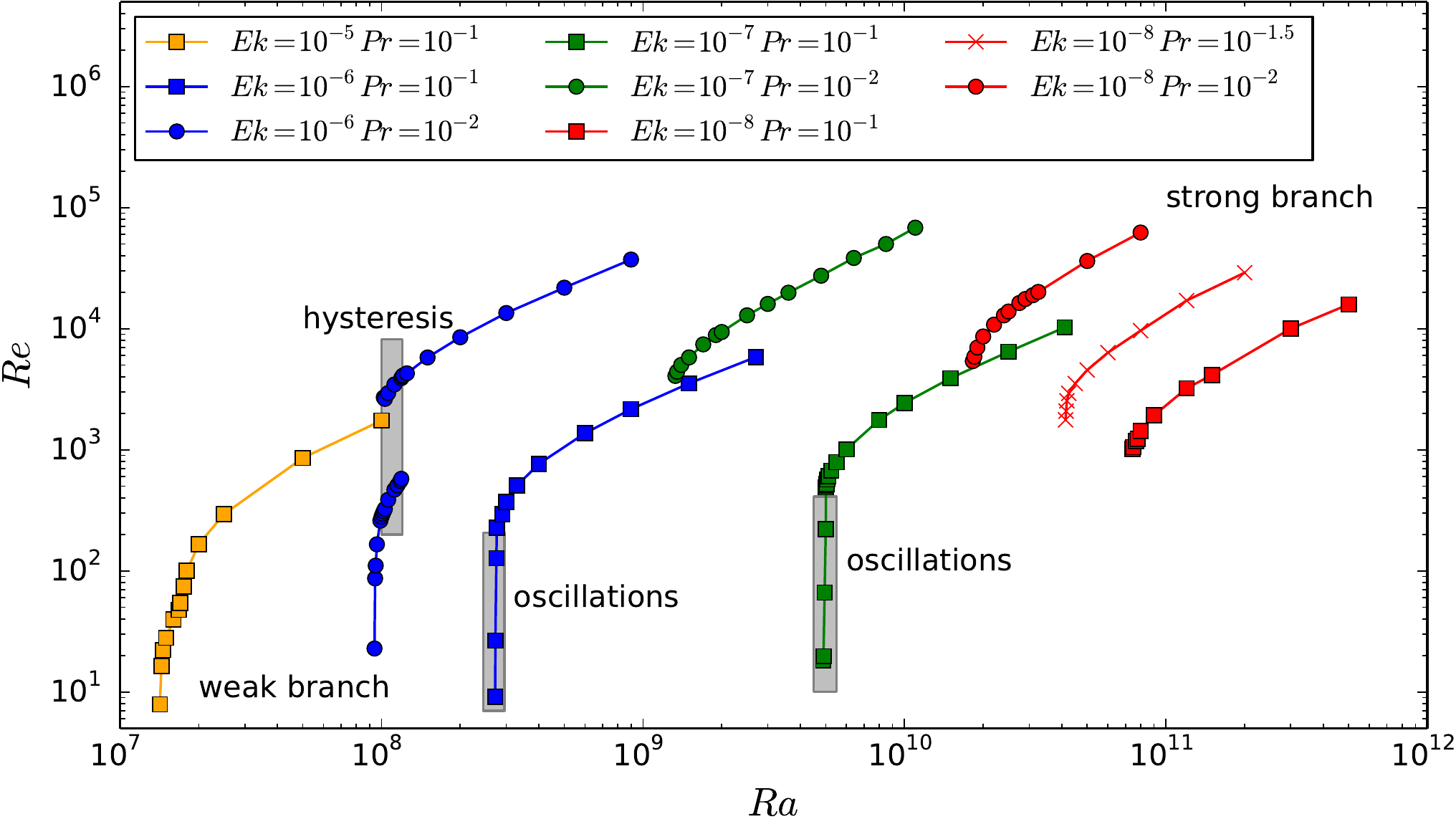}
   \caption{Time-averaged values of the Reynolds number, $\urms$, as a function of the Rayleigh number 
    for different $\Ek$ (identified by different colours) and $\Pran$ (different symbols). 
    The annotations correspond to the different regimes discussed in \S\ref{sec:NL}.}
\label{fig:ReRa}
\end{figure}

To give an overview of the simulations and the results, 
figure~\ref{fig:ReRa} shows the Reynolds number, $\urms$, as a function of the Rayleigh number
for the different Ekman numbers  (identified by different colours) and Prandtl numbers (different symbols).
$\urms$ reaches values up to $10^{5}$ for the lowest Ekman and Prandtl numbers.
Note that the local Rossby number, defined as the rotation period to the convective turnover time, 
is smaller than unity at all lengthscales in all of our simulations, in accordance with our approximation that 
the vortices are columnar at all scales \citep[\eg][]{Nat15}.
At moderate Ekman numbers ($\Ek=10^{-5}$), we recover results previously described
in the literature, such as the sequence of bifurcations between quasi-steady and
time-dependent convection. 
This is discussed in \S\ref{sec:E5}. 
In \S\ref{sec:E8}, we describe a novel branch of convection and the occurrence of subcritical convection at low $\Ek$ ($\Ek=10^{-8}$).  
The bridge between previous results at $\Ek=10^{-5}$ and the new hydrodynamical regime at lower Ekman numbers is made 
at intermediate Ekman numbers ($\Ek=10^{-7}-10^{-6}$), where we observe interesting behaviours: a hysteresis loop
(described in \S\ref{sec:E6}) and nonlinear oscillations (\S\ref{sec:E7}).
These different behaviours are annotated in figure~\ref{fig:ReRa}.

\subsection{Supercritical convection for $\Ek = 10^{-5}$}
\label{sec:E5}

\begin{figure}
\centering
   \subfigure[]{\label{fig:E5_Psia}
   \includegraphics[clip=true,height=4.4cm]{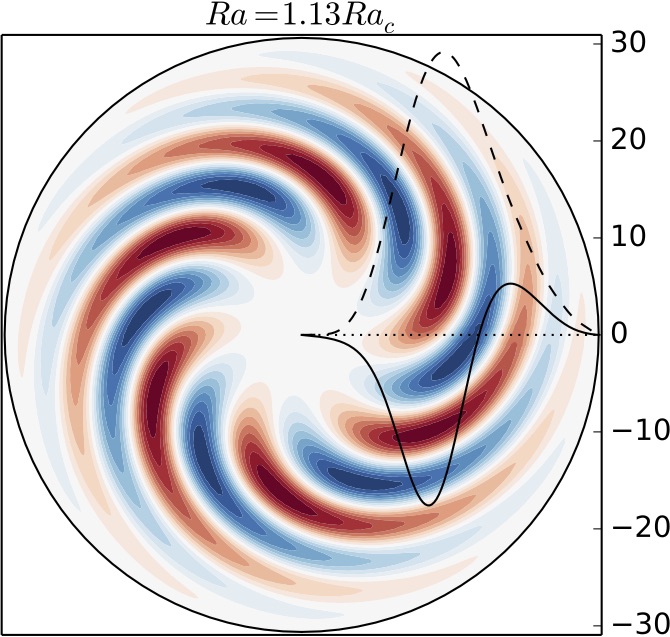}}
    \subfigure[]{\label{fig:E5_Psib}
   \includegraphics[clip=true,height=4.4cm]{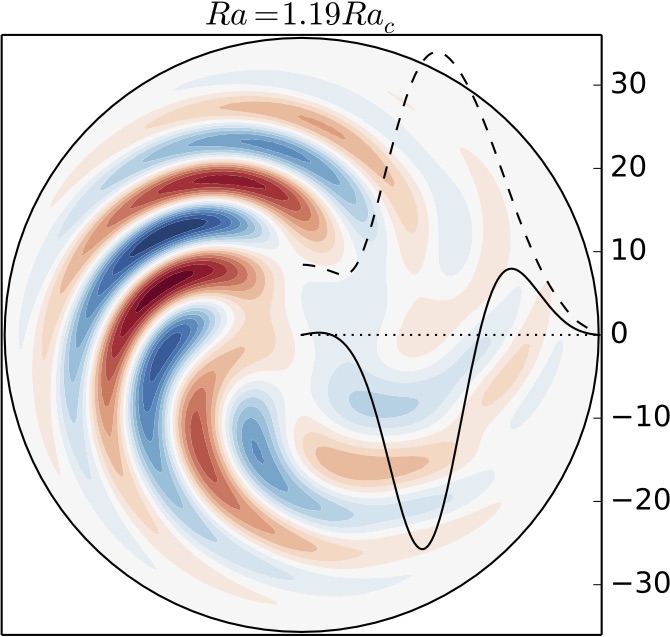}}
   \subfigure[]{\label{fig:E5_Psic}
   \includegraphics[clip=true,height=4.4cm]{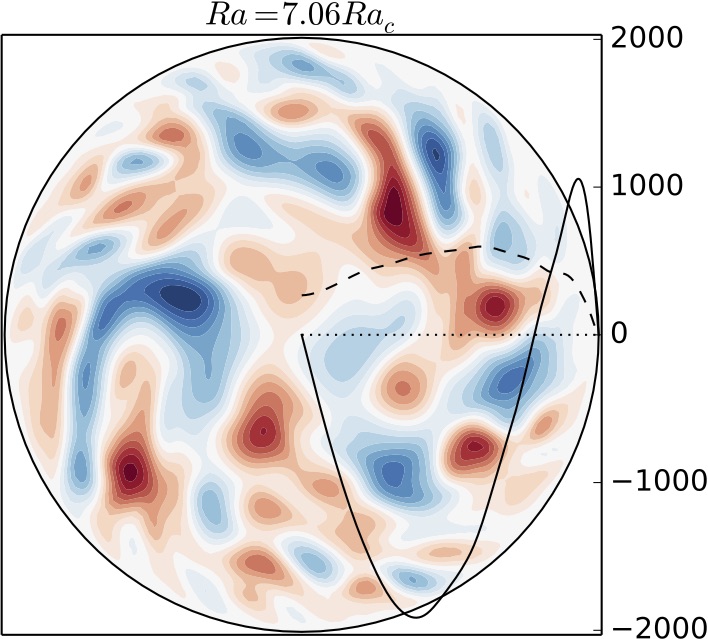}}
   \caption{Snapshots of the streamfunction for $\Ek = 10^{-5}$ and $\Pran = 10^{-1}$ for (a) 
   $\Ra=1.13\Ra_c$, (b) $\Ra=1.19\Ra_c$ and (c) $\Ra=7.06\Ra_c$. 
   The time-averaged radial profiles of the zonal velocity, $\uzon$, (solid line) and the rms radial velocity, $\us$, (dashed line)
   are plotted according to the axis on right-hand side.}
\label{fig:E5_Psi}
\end{figure}

We begin our study with simulations performed at $\Ek=10^{-5}$ and $\Pran=10^{-1}$ in order to link the results obtained with our hybrid QG-3D 
model to previous studies using either 3D or QG models.

For this Ekman number, 
the bifurcation from the basic conduction state is supercritical at $\Ra=\Ra_c$.
For values of $\Ra$ just above $\Ra_c$, the convective flow consists of a thermal Rossby wave, 
a well-known flow pattern in rotating convection \citep[\eg][]{Bus70}. 
Figure~\ref{fig:E5_Psia} shows a snapshot of the streamfunction (colour) for $\Ra=1.13\Ra_c$.
The linear thermal Rossby wave develops around the radius $s\approx 0.5$ and propagates eastward.
For small Prandtl numbers, the thermal Rossby wave has a large tilt in the equatorial cross-section \citep{Zha92}.

\begin{figure}
\centering
   \subfigure[]{\label{fig:E5_urms}
   \includegraphics[clip=true,height=4.7cm]{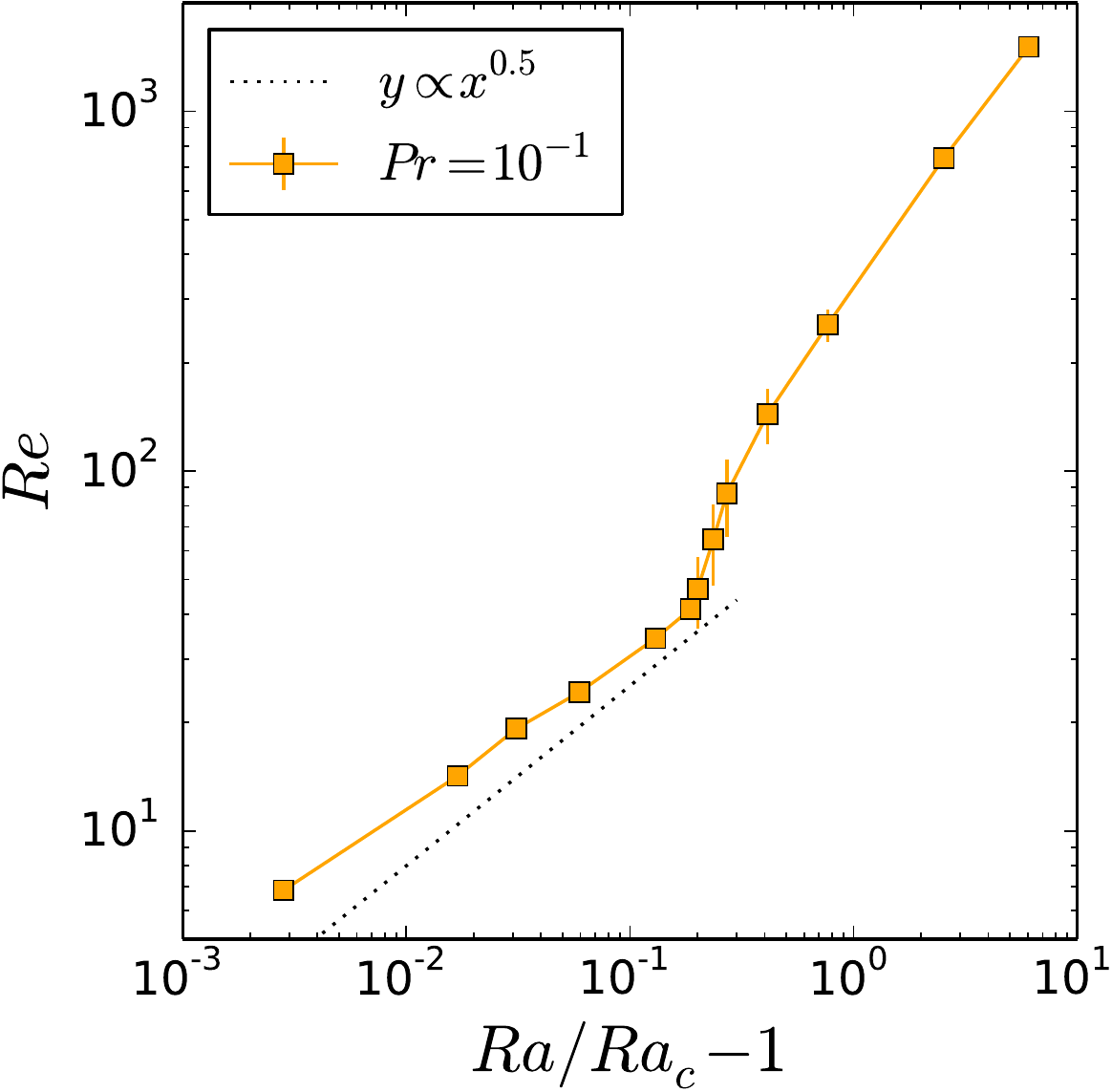}}
    \subfigure[]{\label{fig:E5_Nu}
   \includegraphics[clip=true,height=4.7cm]{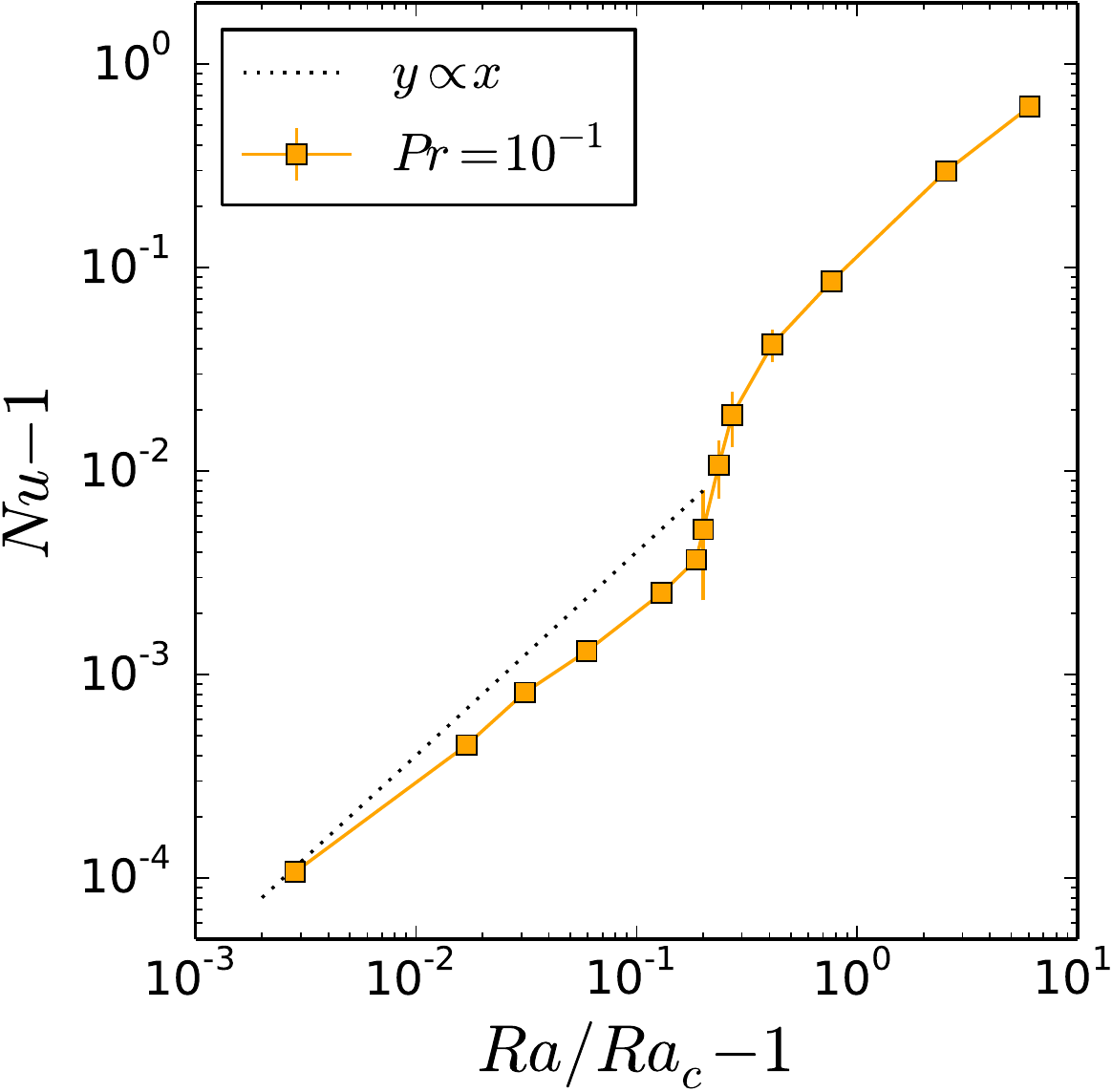}}
   \subfigure[]{\label{fig:E5_uzon}
   \includegraphics[clip=true,height=4.7cm]{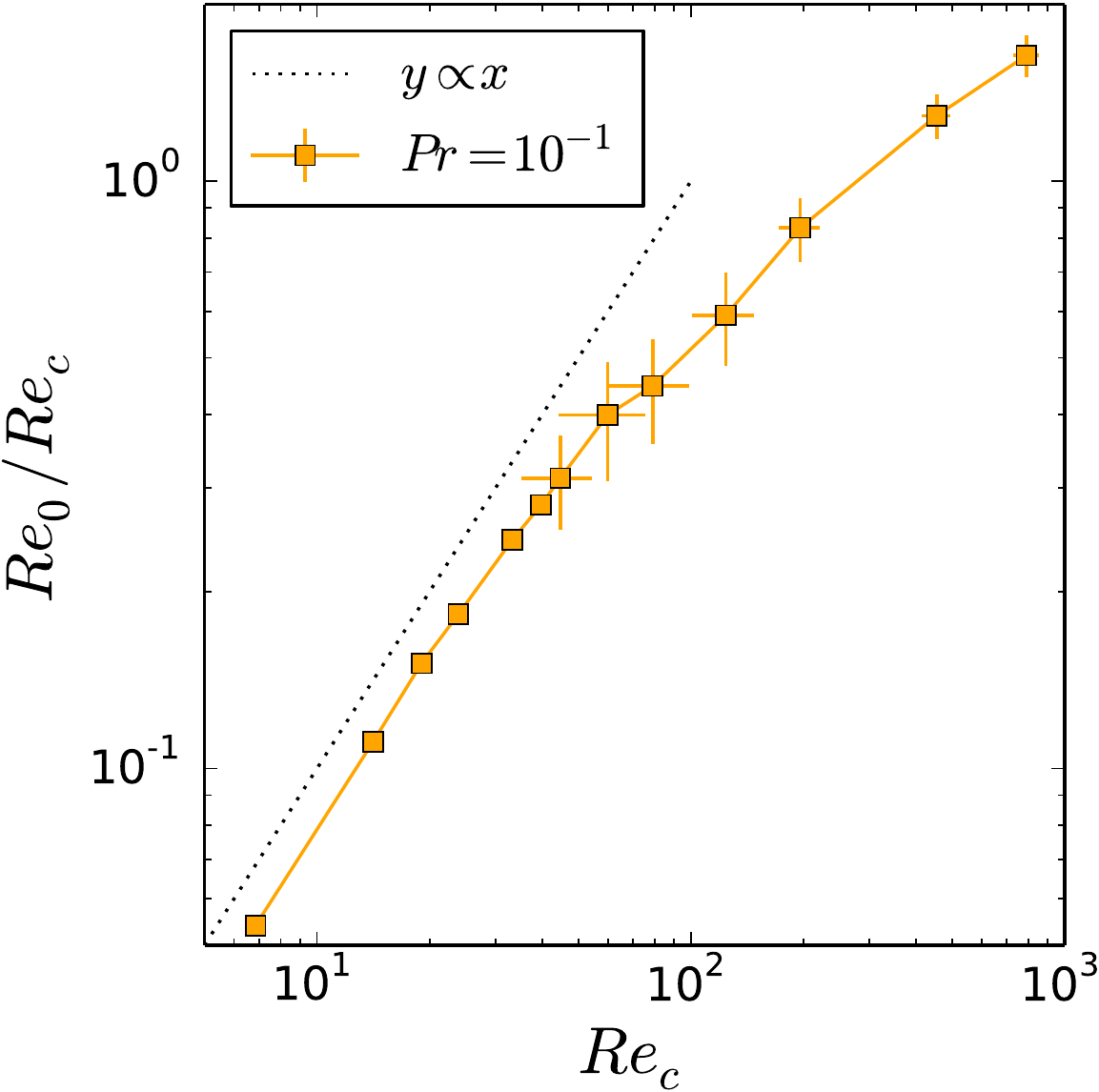}}
   \caption{Evolution of the global properties of convection for $\Ek=10^{-5}$ and $\Pran=10^{-1}$.
   The vertical and horizontal error bars represents the standard deviation in the time series.}
\end{figure}

Figure~\ref{fig:E5_urms} shows the evolution of $\urms$ as a function of the normalised Rayleigh number, $\Ra/\Ra_c-1$.
For $\Ra/\Ra_c<1.2$, the evolution of $\urms$ with $\Ra/\Ra_c-1$ can be approximated by a power law 
$\urms\sim (\Ra/\Ra_c-1)^{0.42}$. 
A sudden transition occurs at $\Ra/\Ra_c=1.2$, after which the slope of the curve becomes steeper.
Around this transition, or kink, the convection changes from quasi-steady to time-dependent.
The sequence of bifurcations that follows the kink has been previously documented in both 3D models
\citep{Zha92b, Sun93, Til97,Gro01, Sim03} and QG models \citep{Schnaubelt1992,Mor04}.
Our hybrid QG model reproduces this sequence: periodic vacillations ($\Ra=1.18\Ra_c$),
chaotic fluctuations with a localisation of the convection ($\Ra=1.19\Ra_c$, figure~\ref{fig:E5_Psib}), 
and bursts of convection ($\Ra=1.21\Ra_c$). 
These various states are attributed to the interactions of the zonal flow and
the axisymmetric temperature with the convective columns \citep{Tee12}.
Interestingly, \citet{Mor04} find that the kink occurs closer to the 
onset of convection as the Ekman number decreases, implying that quasi-steady convection in the form of thermal Rossby waves
is expected to occur within a vanishingly small range of Rayleigh numbers for $\Ek \to 0$. 
For higher Rayleigh number, the time dependence of the convection becomes
irregular. Figure~\ref{fig:E5_Psic} shows the streamfunction for $\Ra=7.06\Ra_c$. The thermal Rossby wave
is no longer visible and the velocity comprises a wide range of lengthscales.

The kink at $\Ra=1.2\Ra_c$ is also clearly visible in the plot of $\Nu$ as a function of $\Ra/\Ra_c-1$ shown in figure~\ref{fig:E5_Nu}. 
Near the onset, \citet{Bus86} predicted that the evolution of the convective heat transport
with the Rayleigh number follows a power law $\Nu\sim\Ra/\Ra_c-1$, which has later been
confirmed by \cite{Gil06} with numerical results for $\Pran>1$. 
For $\Ra<1.2\Ra_c$, our numerical results can be approximated by a power law
of weaker exponent, $0.88$. 
The slow increase of $\Nu$ with the Rayleigh number for $\Pran<1$ was previously observed
by \citet{Gil06}.
After the kink, we have too few numerical points to determine whether 
the points still follow a power law.

In figure~\ref{fig:E5_Psia}, we plot the time-averaged radial profile of the zonal velocity, $\uzon$, for 
$\Ra=1.13\Ra_c$. 
Zonal flows are produced from the nonlinear interactions of the non-axisymmetric velocity components (eq.~(\ref{eq:uzonal_f})). 
The production of zonal flow by thermal Rossby waves has been extensively studied in 
the literature \citep[\eg][]{Bus82,Pla08}. 
In spherical geometry, the divergence
of the Reynolds stresses produces a retrograde zonal jet in the central part and a prograde jet
in the outer part. The correlation of the velocity components is caused by the tilt of the thermal Rossby waves. 
Figure~\ref{fig:E5_Psic} shows $\uzon$ for $\Ra=7.06\Ra_c$. The zonal flow still has a double jet structure with a prograde jet 
on the outer part. The peak velocity of the zonal flow is three times larger than the maximum of the radial profile of the rms radial velocity,
$\us$, also shown in the figure.

Weakly nonlinear analysis predicts that the amplitude 
of the zonal flow scales as the square of the amplitude of the convective velocity \citep{Bus82}.
Figure~\ref{fig:E5_uzon} shows the ratio $\uzrms/\ucrms$ as a function of the convective velocity $\ucrms$.
This ratio varies linearly with $\ucrms$ just above the onset, in agreement with the weakly nonlinear analysis. 
After the transition from quasi-steady convection to time-dependent convection, 
the evolution of $\uzrms/\ucrms$ with $\ucrms$ departs from the weakly nonlinear prediction
and the slope becomes shallower.
The transition occurs when $\uzrms$ becomes comparable to $\ucrms$, in agreement with the idea that  the bifurcation to time-dependent convection is due to the presence of zonal flows.

\subsection{Subcritical convection for $E = 10^{-8}$}
\label{sec:E8}

\begin{figure}
\centering   
	\subfigure[]{\label{fig:t_En_E8P2}
  	\includegraphics[clip=true,width=12cm]{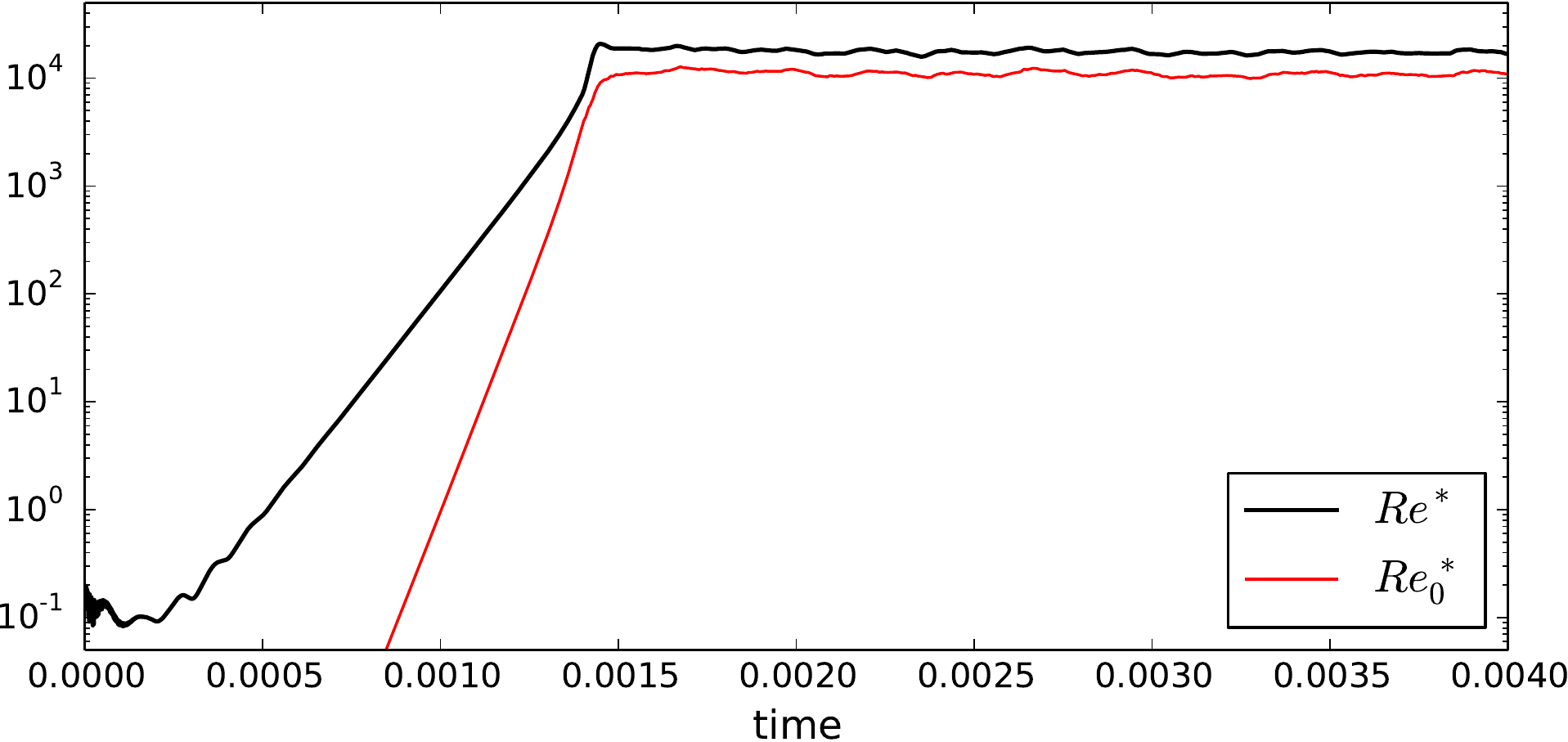}}
	\subfigure[]{\label{fig:Psi_growth_E8P2}
  	\includegraphics[clip=true,height=6.5cm]{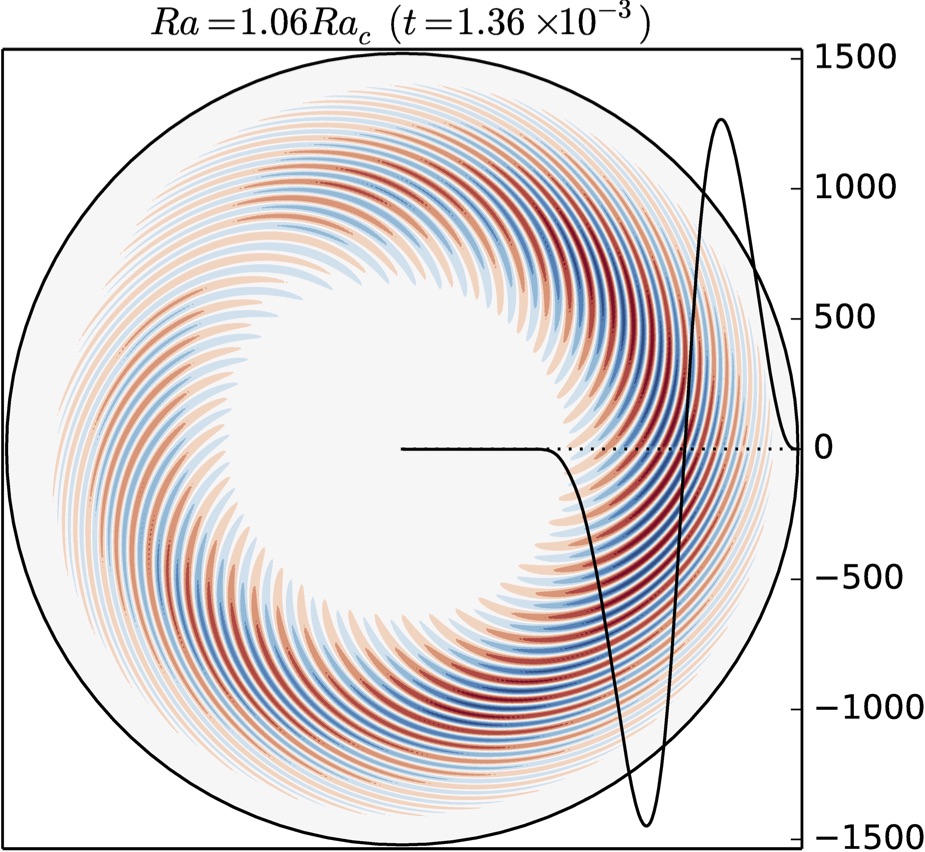}}
	\subfigure[]{\label{fig:Psi_sat_E8P2}
  	\includegraphics[clip=true,height=6.5cm]{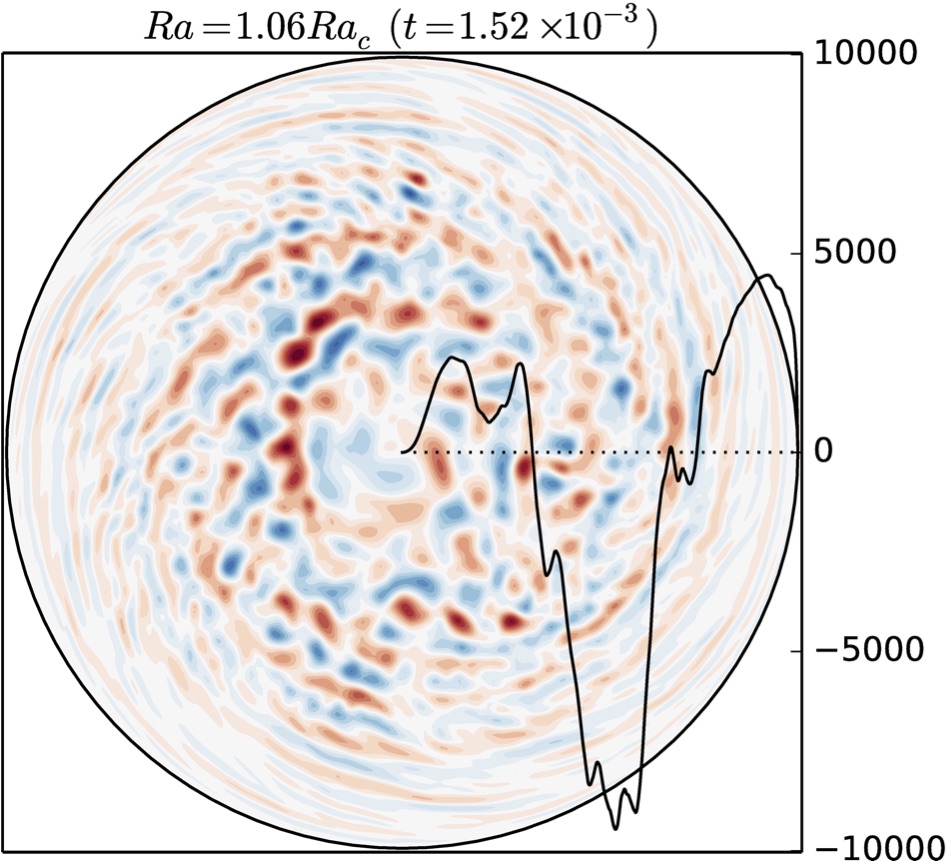}}	
   	\caption{(a) Time series of $\urms^{\ast}$ (thick black line) and  
	$\uzrms^{\ast}$ (red) for $\Ek=10^{-8}$, $\Pran=10^{-2}$ and $\Ra=1.06\Ra_c$. 
	(b)-(c) Streamfunction (colour, snapshot) and radial profiles of the zonal velocity (snapshot, solid line, plotted according to the right-hand side axis)
	during (b) the growing phase at $t=1.36\times10^{-3}$ and (c) the saturated phase at $t=1.52\times10^{-3}$.}
\end{figure}

We now present results obtained at the small Ekman numbers that are out of reach of earlier and recent 3D models \citep[\eg][]{Yad16}.  

Figure~\ref{fig:t_En_E8P2} shows the time series of the Reynolds number, $\urms^{\ast}$, 
and the zonal Reynolds number, $\uzrms^{\ast}$,
for $\Ek=10^{-8}$, $\Pran=10^{-2}$ and $\Ra=1.06\Ra_c$. After a small temperature perturbation is added at $t=0$,
the kinetic energy grows exponentially and eventually saturates around a mean value. 
Despite the proximity to the onset of convection, the values of the Reynolds numbers are large in the saturated phase, 
on average $\urms \approx 15200$ and $\uzrms \approx 9600$.
Figures~\ref{fig:Psi_growth_E8P2} and~\ref{fig:Psi_sat_E8P2} show snapshots of the streamfunction during the growing phase and 
the saturated phase respectively. During the growth, the flow has the distinctive pattern of a thermal Rossby wave. 
The azimuthal modulation of the non-axisymmetric pattern is caused by the interaction
of Rossby waves of different azimuthal wavenumbers. 
A snapshot of the zonal flow is also shown as a solid black line. 
The zonal flow has the familiar double jet structure seen at the onset of convection at $\Ek=10^{-5}$ (figure~\ref{fig:E5_Psia}).
During the saturated phase, the flow is markedly different. The flow is now organised into two regions in the equatorial plane. 
In the inner region ($s<0.6$), the pattern of the thermal Rossby waves is lost and the convective flow is vigorous with a wide range of lengthscales. 
In the outer part ($s>0.6$), the velocity has a smaller amplitude and the characteristic elongated and tilted 
pattern of the Rossby waves is visible.
This type of zonation of the convection was previously
described by \citet{Sum00,Aub03,Miy10} for Rayleigh numbers several times above critical. 
In the outer part, the vortex stretching term due to the slope of the boundaries
is predominant over the other forces, so Rossby waves are easily 
excited there. 
The zonal flow is also modified: its radial structure is more complex with 
three main jets of similar radial widths, where a jet is defined by a zonal velocity of same sign. The innermost and outermost
jets remain prograde while the middle jet is retrograde. 
The outermost prograde zonal jet is driven by the Reynolds stresses from the tilted thermal Rossby waves in the outer region.

\begin{figure}
\centering   
	\subfigure[]{\label{fig:Utot_E8P2}
  	\includegraphics[clip=true,height=7cm]{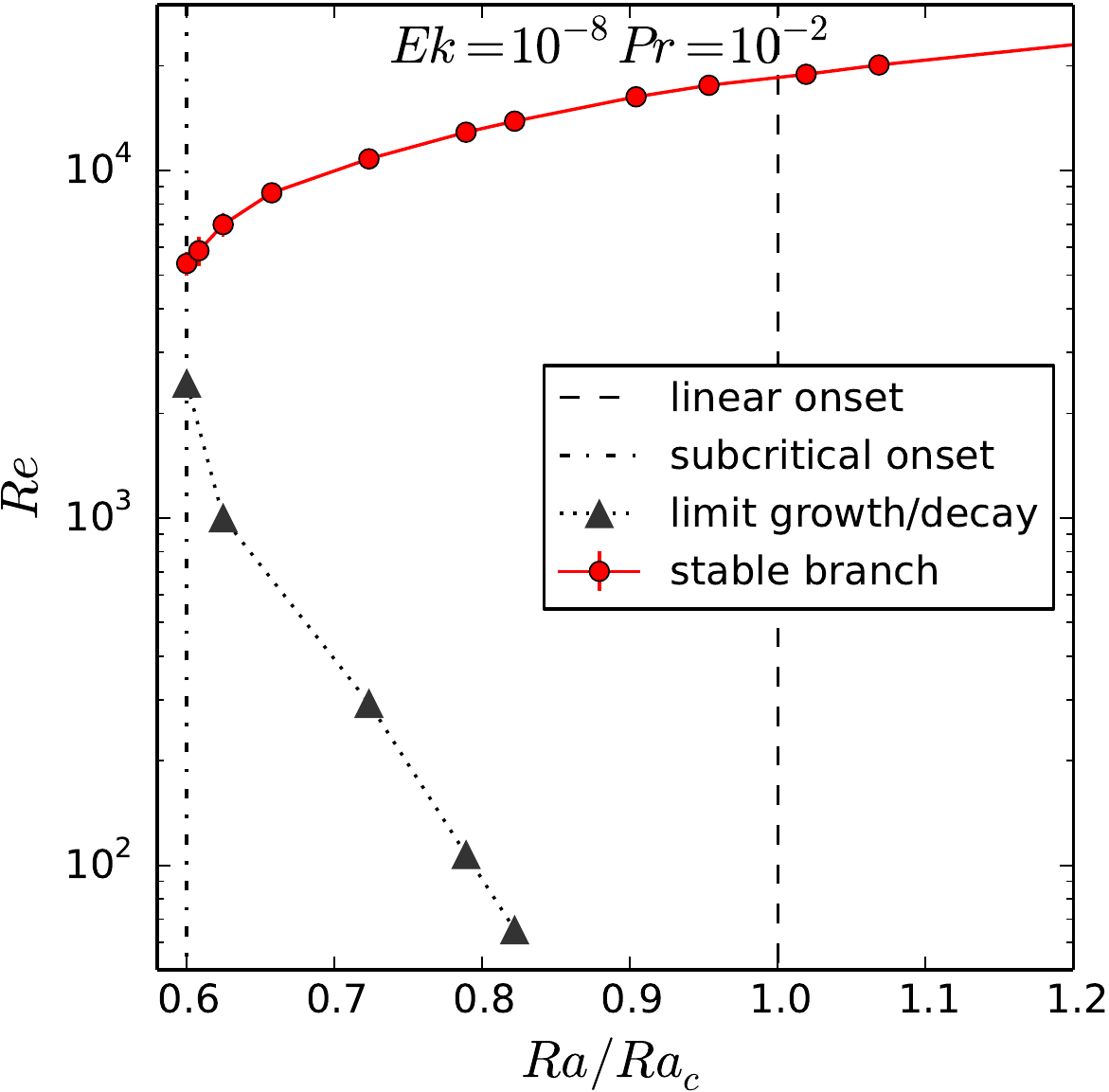}}
		\subfigure[]{\label{fig:Utot_sub}
  	\includegraphics[clip=true,height=7cm]{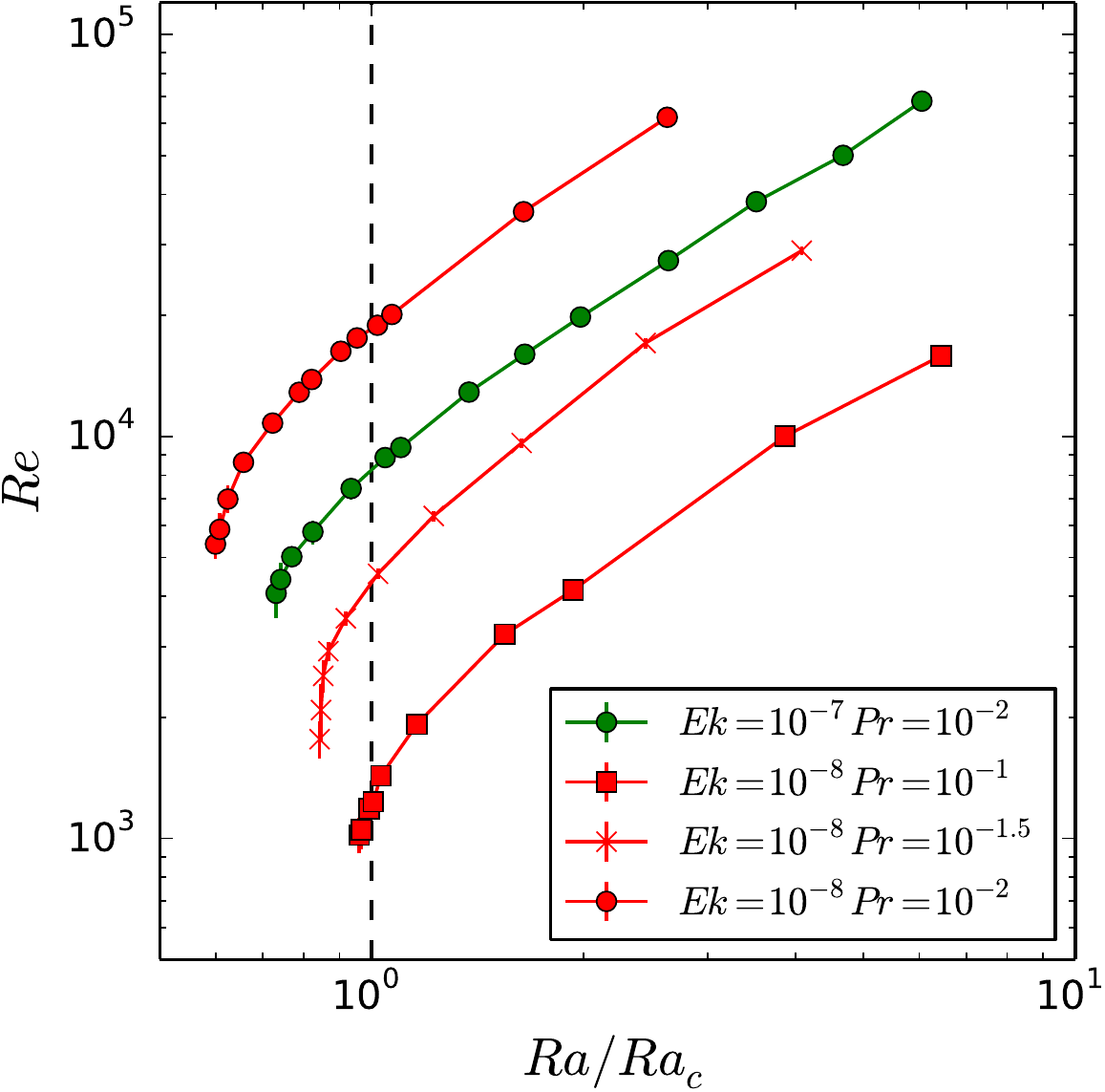}}
   	\caption{(a) $\urms$ as a function of $\Ra/\Ra_c$ for $\Ek=10^{-8}$ and $\Pran=10^{-2}$. 
	The limit between the initial value of ${\urms^{\ast}}$ leading to either the decay or the growth of the solution is plotted as triangles.
	(b) Same as (a) for all the cases of subcritical convection ($\Ek=10^{-8}$, $\Pran\in[10^{-2},10^{-1}]$
	and $\Ek=10^{-7}$, $\Pran=10^{-2}$).}
\end{figure}

\begin{figure}
\centering   
	\subfigure[]{\label{fig:Psi_E8P2Ra1.83}
  	\includegraphics[clip=true,height=6.5cm]{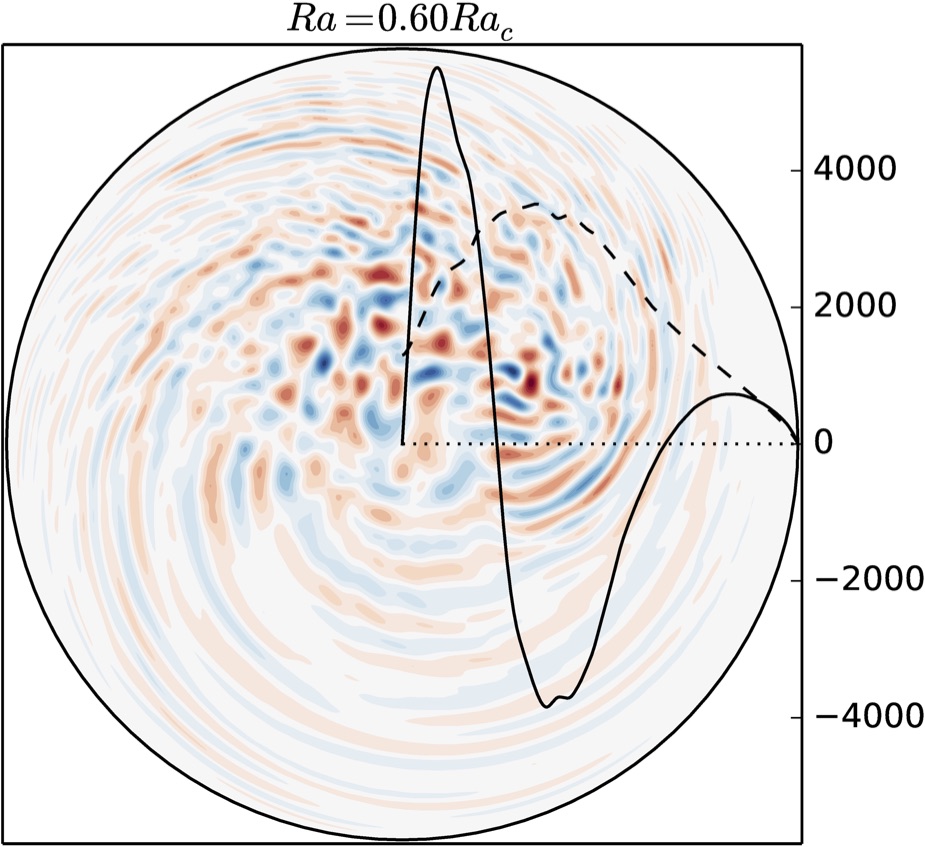}}
	\subfigure[]{\label{fig:Psi_E8P2Ra2}
  	\includegraphics[clip=true,height=6.5cm]{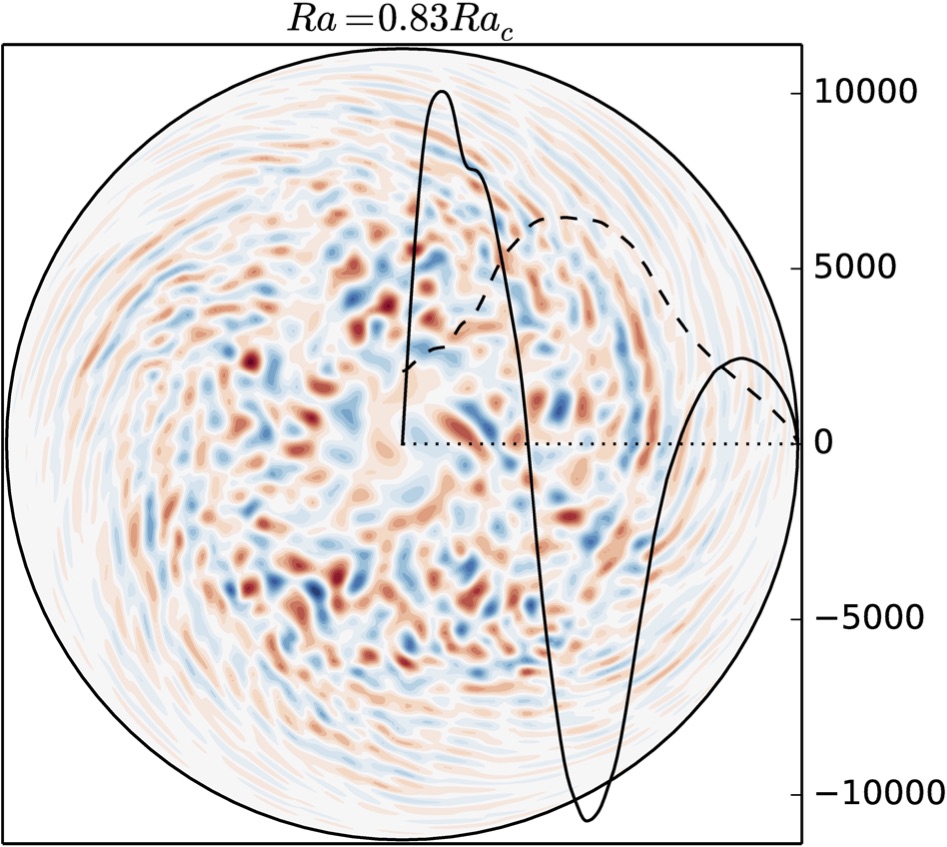}}
   	\caption{Streamfunction (colour, snapshot) and radial profiles of $\us$ (dashed line) 
   	and $\uzon$ (solid line) (time averages) for $\Ek=10^{-8}$, $\Pran=10^{-2}$:
   	(a) $\Ra = \Ra_s=0.60 \Ra_c$ and (b) $\Ra=0.83\Ra_c$.}
\end{figure}

Figure~\ref{fig:Utot_E8P2} shows the values of $\urms$ when the Rayleigh number is varied around the 
onset of convection for $\Ek = 10^{-8}$ and $\Pran=10^{-2}$.
As previously noted in figure~\ref{fig:t_En_E8P2},
$\urms$ becomes large (of the order of $10^4$) when $\Ra$ is only just slightly increased above 
$\Ra_c$. 
Remarkably, when the Rayleigh number is decreased during a simulation 
from $\Ra\geq \Ra_c$ to $\Ra<\Ra_c$, the convection does not decay and
remains vigorous up to a value of the Rayleigh number, $\Ra_s$, where the convection abruptly
shuts down. At $\Ra=\Ra_s$, the value of $\urms$ is discontinuous: 
for Rayleigh numbers above $\Ra_s$, we have $\urms>5000$, whereas for all $\Ra$ below
$\Ra_s$, $\urms=0$.
This behaviour ($\Ra_s<\Ra_c$) is typical of a subcritical onset of nonlinear convection.
For $\Ek=10^{-8}$ and $\Pran=10^{-2}$, we find that $\Ra_s/\Ra_c = 0.60$.
Figures~\ref{fig:Psi_E8P2Ra1.83}-\ref{fig:Psi_E8P2Ra2} show snapshots of the streamfunction for 
$\Pran=10^{-2}$ and two Rayleigh numbers on the subcritical branch, $\Ra/\Ra_c =0.60$ and $0.83$. 
For $\Ra=\Ra_s = 0.60\Ra_c$, the convective flow is similar to the saturated case described 
previously in figure~\ref{fig:Psi_sat_E8P2}.
The main difference is that the amplitude of the flow in the inner convective region is 
concentrated in a localised structure, that drifts in the azimuthal direction with time.
The localisation of the convection might indicate that the system struggles to 
maintain convection for this Rayleigh number. However the kinetic energy of the convective and zonal flows do not decay
during the entire time integration of the simulation, which corresponds to  $200$ convective turnover timescales 
(where one turnover timescale is calculated as $1/\ucrms$), or equivalently, $2$ global thermal diffusion timescales.
For larger Rayleigh numbers, 
the localisation in the inner convective region disappears
and the convection fills all longitudes, as observed for $\Ra=0.83\Ra_c$ in figure~\ref{fig:Psi_E8P2Ra2}. 
The radial profile $\us$ is plotted in figures~\ref{fig:Psi_E8P2Ra1.83}-\ref{fig:Psi_E8P2Ra2}.
The rms radial velocity peaks in the inner convective region, and then, monotonically 
decreases in the outer Rossby wave region. The boundary between the convective inner region 
and the outer Rossby wave region moves outwards when $\Ra$ increases.  
The radial profile of the zonal flow, $\uzon$, is also plotted in 
figures~\ref{fig:Psi_E8P2Ra1.83}-\ref{fig:Psi_E8P2Ra2}. As described in the saturated case of figure~\ref{fig:Psi_sat_E8P2},
the zonal flow displays three jets in both cases. The peak of the middle retrograde jet roughly coincides with the maximum of $\us$.  
The zonal flows represent a significant
portion of the total flow with the amplitude of $\uzrms$ increasing from $0.5\ucrms$ to $1.3\ucrms$ when $\Ra/\Ra_c$ varies from $0.60$ to $3$.
No significant change in the flow is observed when the Rayleigh number is varied across the linear critical value $\Ra_c$,
and all of the output parameters evolve continuously, similarly to $\ucrms$ in figure~\ref{fig:Utot_E8P2}. 
No continuous branch of convection is found at $\Ra=\Ra_c$ when the convection is started from a small perturbation as described for $\Ek=10^{-5}$.

Within the interval $\Ra_s\leq\Ra<\Ra_c$, the system has two stable solutions: one is located on the subcritical branch
shown in red in figure~\ref{fig:Utot_E8P2} and is found by using an initial condition with a large enough amplitude; the other solution corresponds to $\urms=0$ 
and is found by starting the simulation from a small perturbation.
In order to quantify how large the finite amplitude of the initial condition must be to access the subcritical branch,
we perform a series of simulations where the initial condition is taken from a snapshot of the saturated solution on the subcritical branch for given parameters
and we initially divide the amplitude of the velocity by a given factor.   
By a trial and error procedure, we determine the factor that separates 
a solution where the kinetic energy decays to zero from a solution where the kinetic energy 
grows back to the value on the subcritical branch. 
The limit between the values of $\urms^{\ast}$ of the initial condition leading to either
the decaying solution or the growing solution are shown in figure~\ref{fig:Utot_E8P2}. 
As expected, we find that for values of $\Ra$ just above $\Ra_s$,
the finite amplitude of the initial condition must be much larger than for values of $\Ra$ just below $\Ra_c$.

\begin{table}
  \begin{center}
\def~{\hphantom{0}}
  \begin{tabular}{lcccc}
      $\Ek$  & $10^{-8}$   &   $10^{-8}$ & $10^{-8}$ & $10^{-7}$ \\[3pt]
      \hline
       $\Pran$   & $10^{-1}$ & $10^{-1.5}$ & $10^{-2}$ & $10^{-2}$\\
       $\Ra_s/\Ra_c$  & 0.96 & 0.84 & 0.60 & 0.73\\
  \end{tabular}
  \caption{Values of the ratio of Rayleigh number at the nonlinear onset of convection to the critical Rayleigh number at the linear onset in the subcritical cases.}
  \label{tab:sub}
  \end{center}
\end{table}

By varying the Prandtl number between $[10^{-2},10^{-1}]$ and the Ekman number between $[10^{-8},10^{-5}]$, 
we find that subcritical convection occurs for $\Ek=10^{-8}$ at $\Pran\leq 10^{-1}$
and $\Ek=10^{-7}$ at $\Pran \leq 10^{-2}$.  Table~\ref{tab:sub} gives the values of $\Ra_s/\Ra_c$ for the subcritical cases and figure~\ref{fig:Utot_sub} 
shows $\urms$ as a function of $\Ra/\Ra_c$ for these cases.
Note that long time integrations are required to determine $\Ra_s$. The values of $\Ra_s$ are obtained from simulations where the convection
does not decay for the entire duration of the calculation, which is at least $200$ convective turnover timescales.
We find that the value of $\urms$ at $\Ra=\Ra_s$ is of the order of or larger than $10^3$. 
The ratio $\Ra_s/\Ra_c$ decreases when $\Ek$ decreases, and also when $\Pran$ decreases. 
Consequently, our results suggest that the subcritical behaviour is amplified by low 
Ekman and Prandtl numbers.

The subcritical behaviour of convection in internal heating models in the limit of small $\Ek$ was anticipated from the work of \citet{Sow77}, who found weakly nonlinear solutions close to the critical Rayleigh number at the linear onset predicted by the local theory of \citet{Bus70},
which is significantly smaller than the true global value \citep{Jon00}.
These nonlinear solutions are maintained in the subcritical domain because the zonal shear produced by nonlinear effects opposes 
the process of phase mixing, which is responsible for the decay of the linear solution.
Subcritical convection for small Ekman numbers
was also predicted by \cite{Pla08} from a weakly nonlinear analysis of 
a quasi-geostrophic model with internal heating. \citeauthor{Pla08} found that 
the bifurcation to the thermal Rossby waves changes from supercritical to subcritical
for $\Ek\leq 6.7\times 10^{-7}$ (in our scaling) at $\Pran=1$. 
Their analysis demonstrates that the nonlinearities in the temperature equation are responsible for the transition. 
These nonlinearities are produced by
the interactions of the zonal flow with the temperature perturbation of the wave and by
the interactions of the axisymmetric temperature with the velocity of the wave.
The difference between our numerical results and the weakly nonlinear analysis of \citeauthor{Pla08},
which predicts subcriticality for larger $\Ek$ at $\Pran=1$, is possibly due to their treatment of the 
temperature in 2D.
As far as we know, the numerical results of the present paper are the first to demonstrate the existence of
subcritical rotating convection in a fully nonlinear model in spherical geometry.
The requirements of small Ekman numbers 
and small Prandtl numbers (at least for $\Ek\in[10^{-8},10^{-7}]$)
can explain why subcriticality has not been observed in previous numerical models
(\eg by \citet{Mor04} with QG simulations at $\Ek=2\times 10^{-7}$ and $\Pran=1$).   
Although subcriticality in the fully 3D system has been predicted from the work of \citet{Sow77} for small $\Ek$,
its presence in our results might yet be a consequence of our quasi-geostrophic approximation. 
Future 3D simulations will be needed to determine whether this is the case.

\subsection{A hysteresis loop for $E = 10^{-6}$ and $Pr=10^{-2}$}
\label{sec:E6}

\begin{figure}
\centering  
   \subfigure[]{\label{fig:Re_hyst} 
   \includegraphics[clip=true,width=7cm]{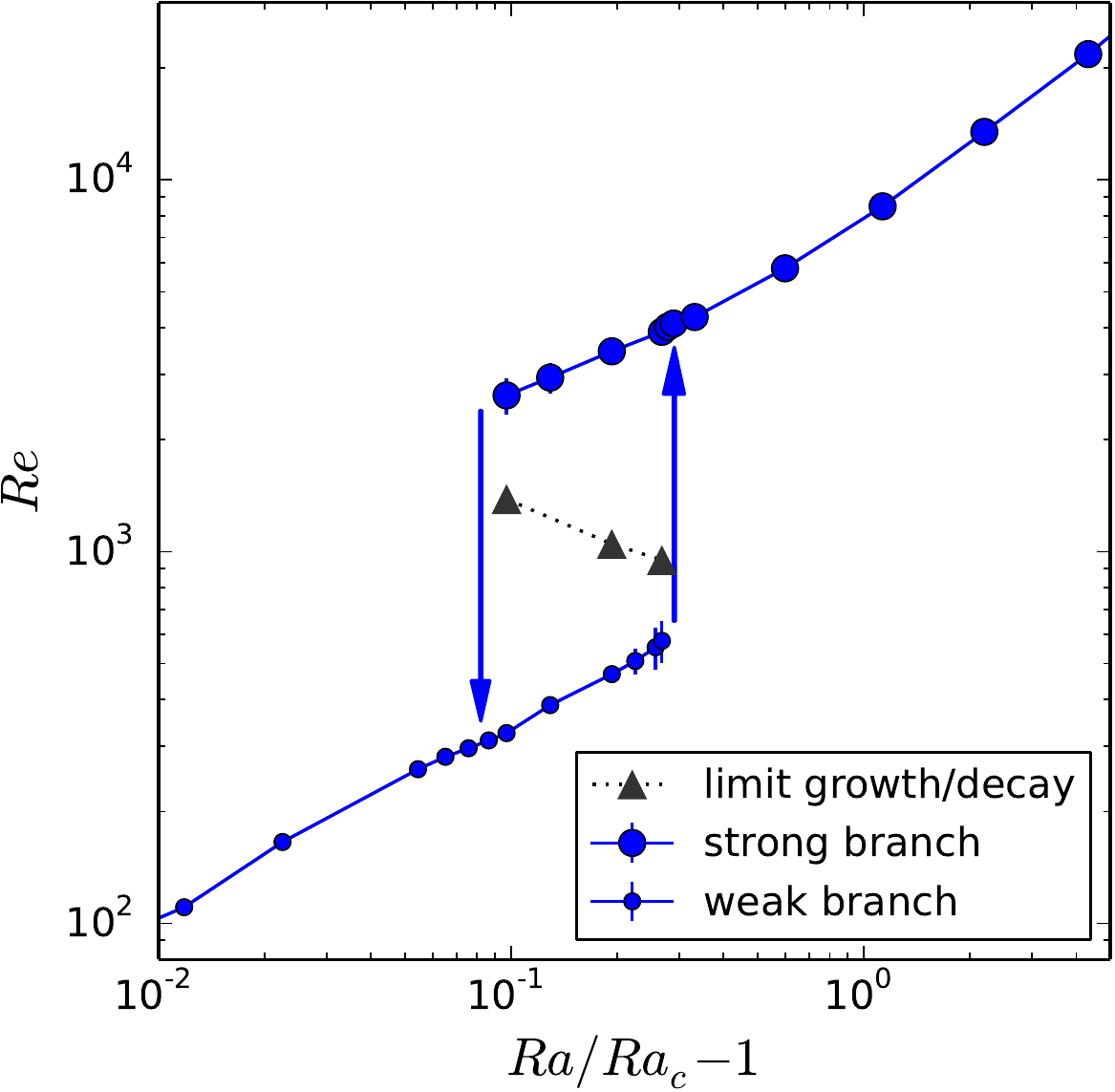}}
   \subfigure[]{\label{fig:Nu_hyst} 
   \includegraphics[clip=true,width=7cm]{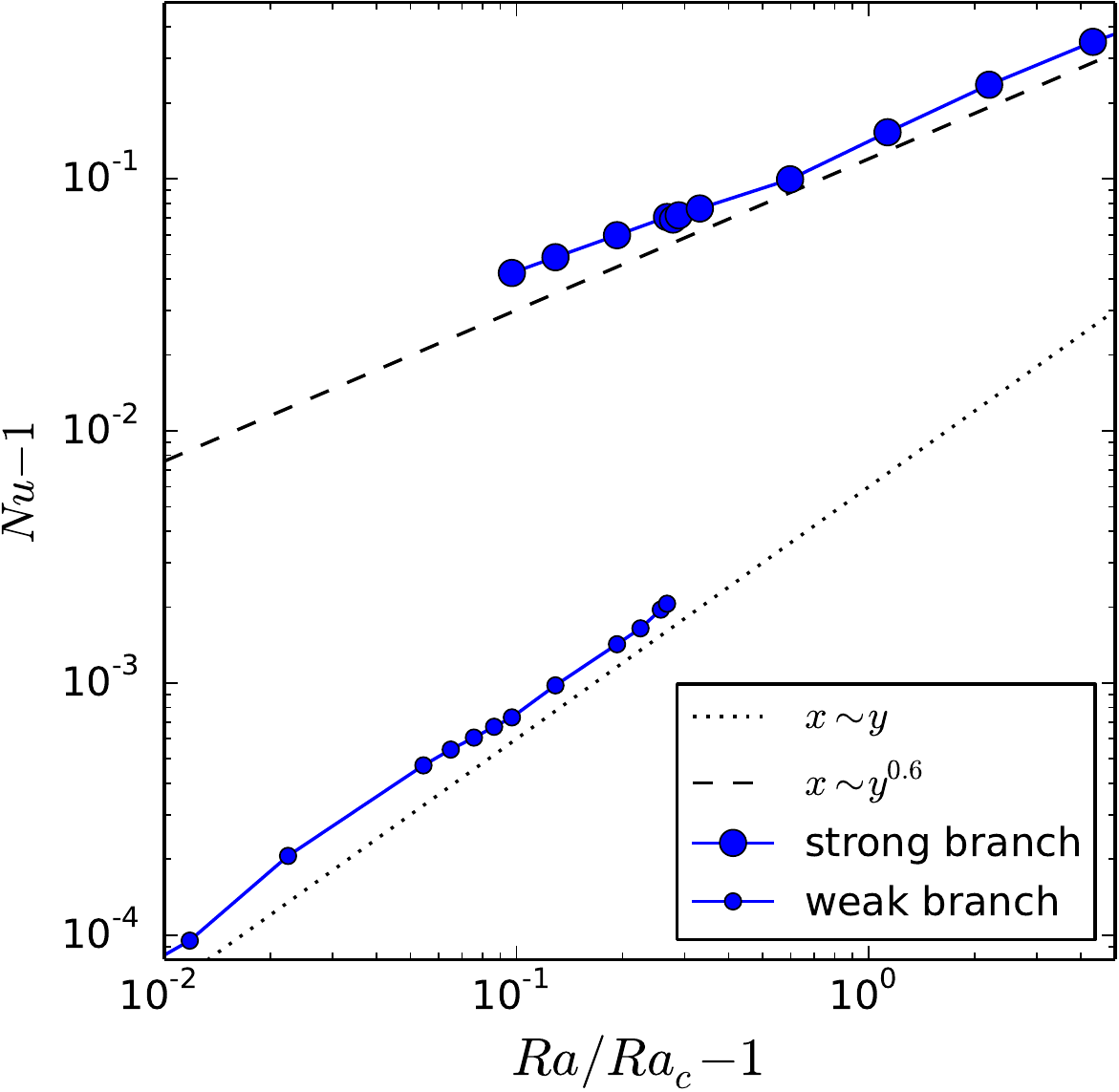}}
   \caption{Evolution of (a) $\urms$ and (b) $\Nu-1$ as a function of the Rayleigh number 
   for $E = 10^{-6}$ and $Pr=10^{-2}$.}
   \label{fig:hyst}
\end{figure}

For $\Ek=10^{-6}$, the bifurcation at the onset of convection is supercritical for all
the Prandtl numbers studied here ($\Pran\in[10^{-2},10^{-1}]$).
For $Pr=10^{-2}$, we observe an interesting behaviour near the onset: two stable branches
of convection co-exist within a limited range of Rayleigh numbers. 
This behaviour can be seen in the evolution of $\urms$ and $\Nu-1$ with $\Ra/\Ra_c-1$ in 
figure~\ref{fig:hyst}. 
On the upper branch, $\urms$ is one order of magnitude larger than on the lower branch.
We thus refer to the upper branch as strong and the lower branch as weak.  
The two branches are stable and co-exist for Rayleigh numbers between $1.09\leq \Ra/\Ra_c\leq1.27$. 
The co-existence of two stable branches for a range of $\Ra$ leads to a hysteresis loop. 
The convection selects the weak or the strong branch depending on the
dynamical history of the system. For instance, the system remains on the strong branch if the 
initial condition is taken from a simulation with $\Ra>1.27\Ra_c$ and $\Ra$ is then decreased down to $1.09\Ra_c$.
For $\Ra>1.27\Ra_c$, only the strong branch exists according to our calculations.
When the system is located on the weak branch, an increase of the Rayleigh number above $1.27\Ra_c$ 
therefore leads the system to jump to the strong branch.

\begin{figure}
\centering   
   \includegraphics[clip=true,width=12cm]{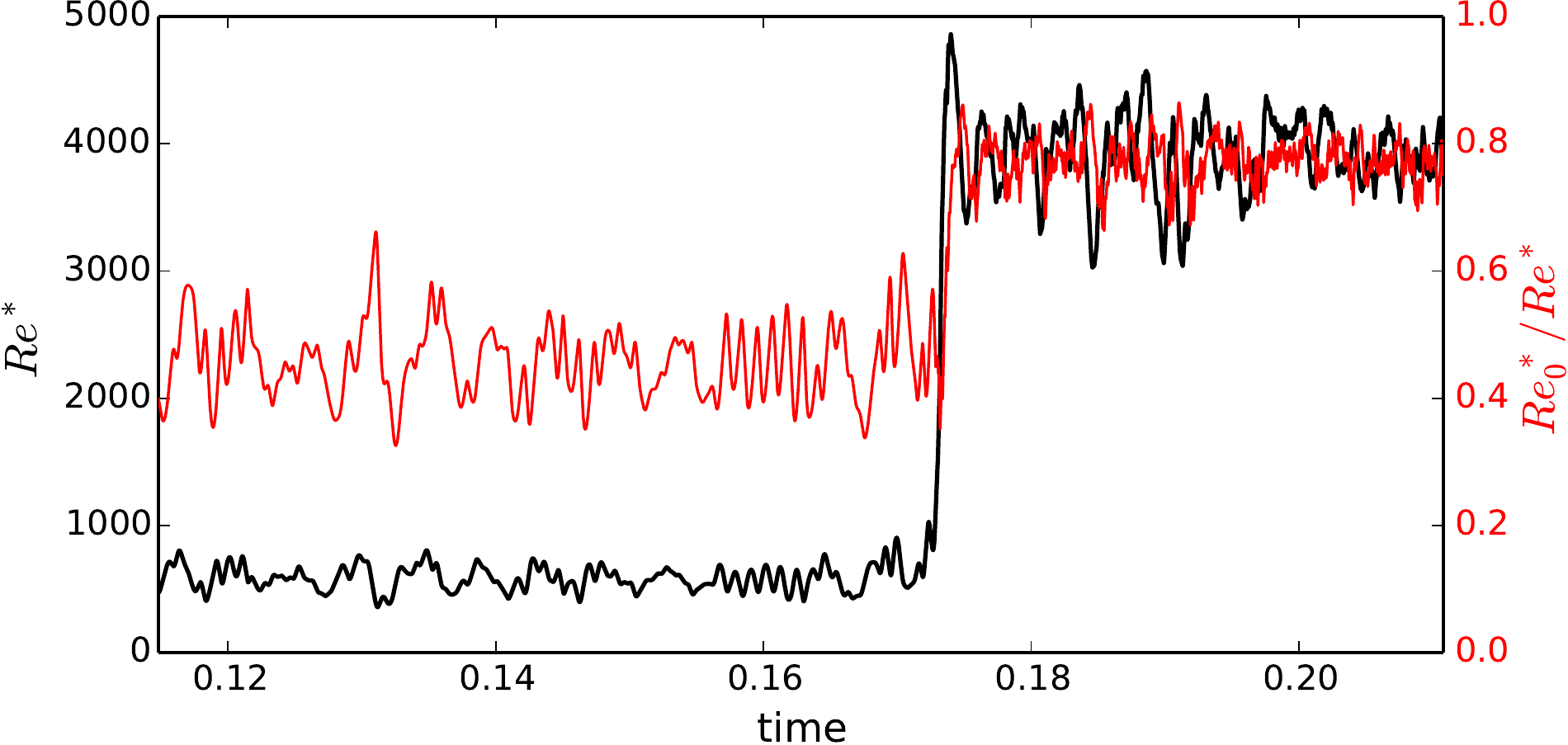}
   \caption{Time series of $\urms^{\ast}$ (thick black line, left axis) and  $\uzrms^{\ast}/\urms^{\ast}$
   (red, right axis) for a simulation near the edge of the hysteresis loop $E = 10^{-6}$, $Pr=10^{-2}$ and $\Ra=1.27\Ra_c$.}
   \label{fig:jump_Ek}
\end{figure}

In order to quantify the amplitude of the velocity required to jump from one branch to the other,
we follow the same numerical procedure as described in \S\ref{sec:E8}.  
As an initial condition, we use a snapshot of a solution on the weak branch that we multiply
by a given factor and determine
the initial amplitude of the flow leading to either a return to the weak branch or
a growth to the strong branch. The triangles in figure~\ref{fig:Re_hyst}
indicate the limit between the initial value of the rms velocity leading to
the growth or the decay of the energy. As expected a larger amplitude of the initial velocity
is required to jump to the strong branch for  $\Ra=1.09\Ra_c$ than $\Ra=1.27\Ra_c$.
In fact, for $\Ra=1.27\Ra_c$, we observe that, when the system is on the weak branch, 
a fluctuation of the kinetic energy can 
lead the system to jump to the strong branch regime. Figure~\ref{fig:jump_Ek}
shows this jump in a time series of the Reynolds number, $\urms^{\ast}$.
The two branches are well separated as there is 
no slow variation on the weak branch towards the strong 
branch, but only a sudden jump that lasts less than $3$ convective turnover timescales.
The ratio of zonal to total Reynolds numbers 
(plotted according to the right axis) is approximately $0.45$ on the weak branch and increases to approximately $0.8$ after the jump to the strong branch.

\begin{figure}
\centering   
  \subfigure[]{\label{fig:Psi_E6P2_weak}
  \includegraphics[clip=true,height=6.5cm]{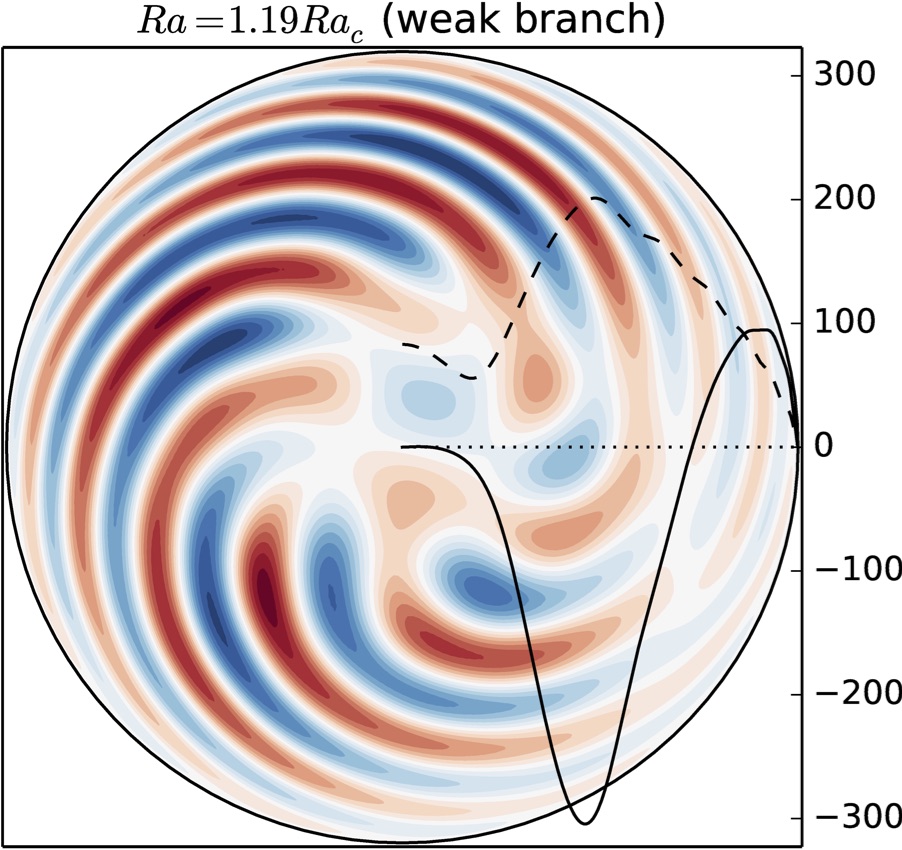}}
  \subfigure[]{\label{fig:Psi_E6P2_strong}
  \includegraphics[clip=true,height=6.5cm]{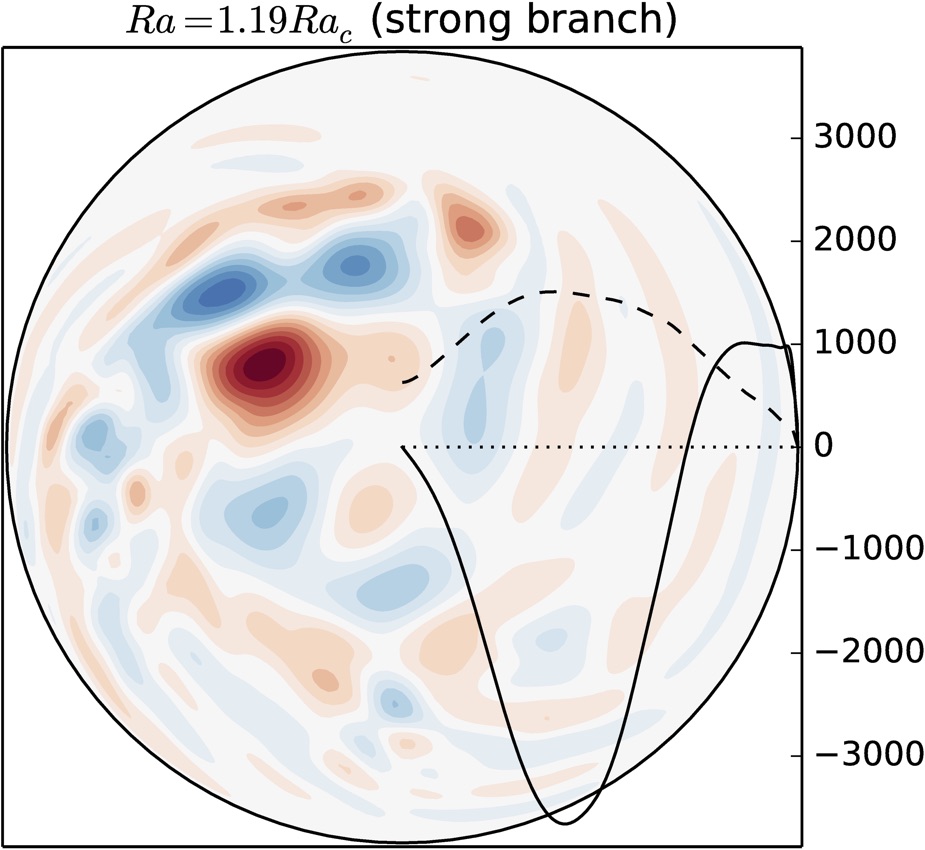}}
  \caption{Streamfunction (colour), $\uzon$ (solid line) and $\us$ (dashed line) for 
  $\Ek = 10^{-6}$, $\Pran=10^{-2}$ and $\Ra=1.19\Ra_c$ on (a) the weak branch and (b) the strong branch.}
\end{figure}

Typical convective flows on the weak and strong branches are shown in figures~\ref{fig:Psi_E6P2_weak}-\ref{fig:Psi_E6P2_strong}
for the same Rayleigh number $\Ra=1.19\Ra_c$. 
On the weak branch, the flow consists in thermal Rossby waves. The zonal flow has a double jet structure and 
its amplitude is comparable to the amplitude of the radial flow. 
The characteristics of this flow are similar to the case described near the onset of convection 
for $\Ek=10^{-5}$ (\S\ref{sec:E5}). 
On the strong branch, the thermal Rossby waves are not visible anymore and a wide range of lengthscales are observed. The zonal flow has a larger amplitude than the radial velocity. 
Not only do the solutions look different on the two branches, the global properties of the convection
also follow different scaling laws with the input parameters.
The evolution of $\Nu$ with the Rayleigh number (figure~\ref{fig:Nu_hyst}) is less steep on 
the strong branch compared with
the weak branch. On the weak branch we find that the points approximately follow a power
law $\Nu\sim(\Ra/\Ra_c-1)^{\alpha}$ with $\alpha=1$, whereas on the strong branch $\alpha=0.6$.

\subsection{Non-linear oscillations for $E = 10^{-7}$ and $Pr=10^{-1}$}
\label{sec:E7}

\begin{figure}
\centering   
 \includegraphics[clip=true,width=12cm]{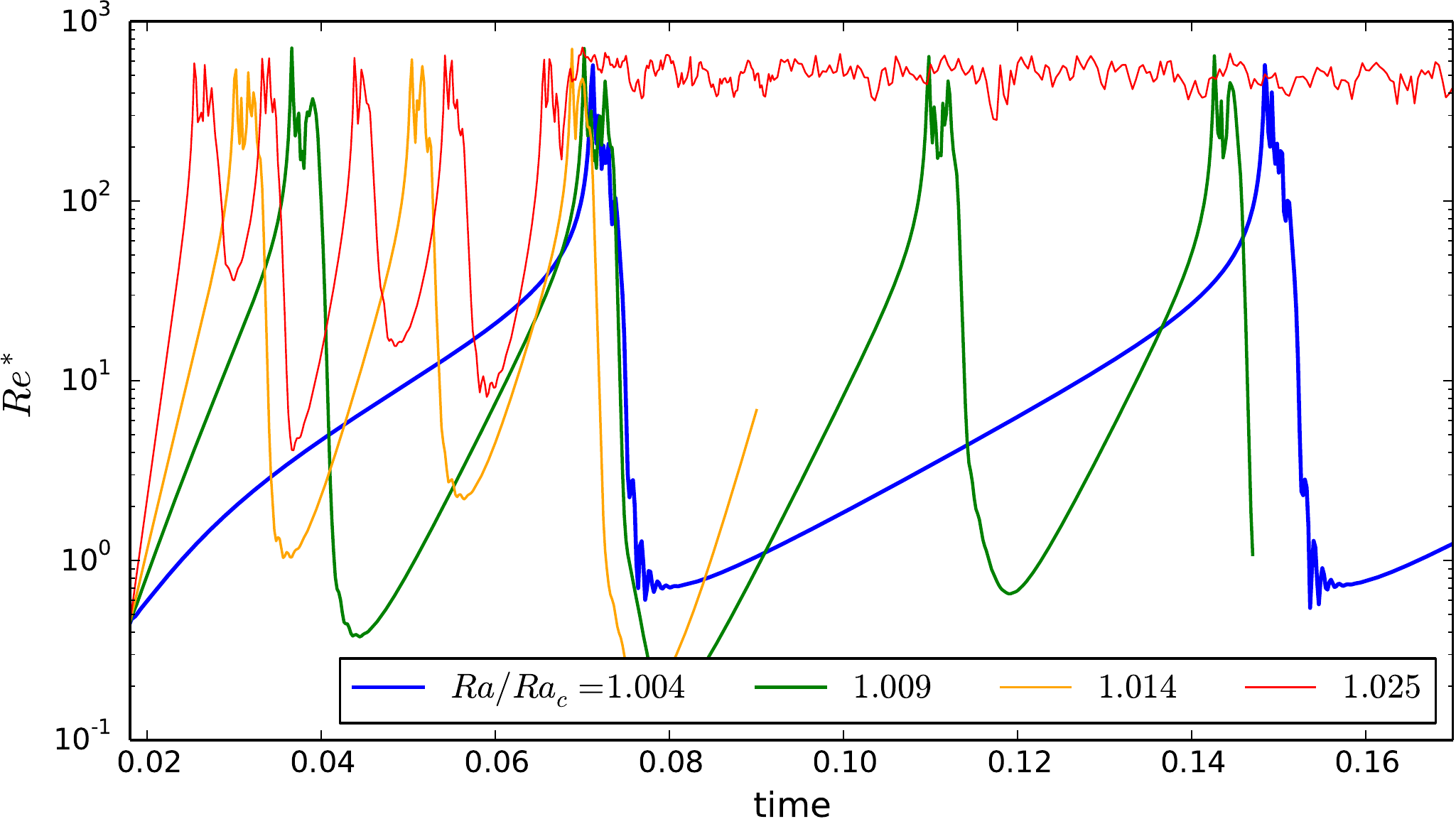}    
   \caption{Time series of the Reynolds number, $\urms^{\ast}$, for $E = 10^{-7}$ and $Pr=10^{-1}$ and Rayleigh numbers near the onset.}
\label{fig:t_En_int}
\end{figure}

For $E = 10^{-7}$ and $Pr=10^{-1}$, we observe an interesting behaviour between the linear onset of convection and the
saturated branch of convection, for values of $\Ra$ between $\Ra_c$ and $1.025\Ra_c$. 
This behaviour is best observed in the time series of $\urms^{\ast}$ shown in figure~\ref{fig:t_En_int}
for Rayleigh numbers in this interval. $\urms^{\ast}$ first grows at a rate that depends on $\Ra$ and 
then reaches a value of approximately $400$.  After a few fluctuations around this mean value, $\urms^{\ast}$ decreases rapidly. 
The system subsequently enters a new growing phase and the oscillation repeats itself. 
The duration of the oscillation depends on the growth rate, and so, larger Rayleigh numbers have shorter period.
For $\Ra=1.025\Ra_c$, the system eventually stays on the saturated branch after a few oscillations, after which it never goes back to the oscillating phase.

\begin{figure}
\centering   
   \subfigure[]{\label{fig:phase_int}
   \includegraphics[clip=true,width=7cm]{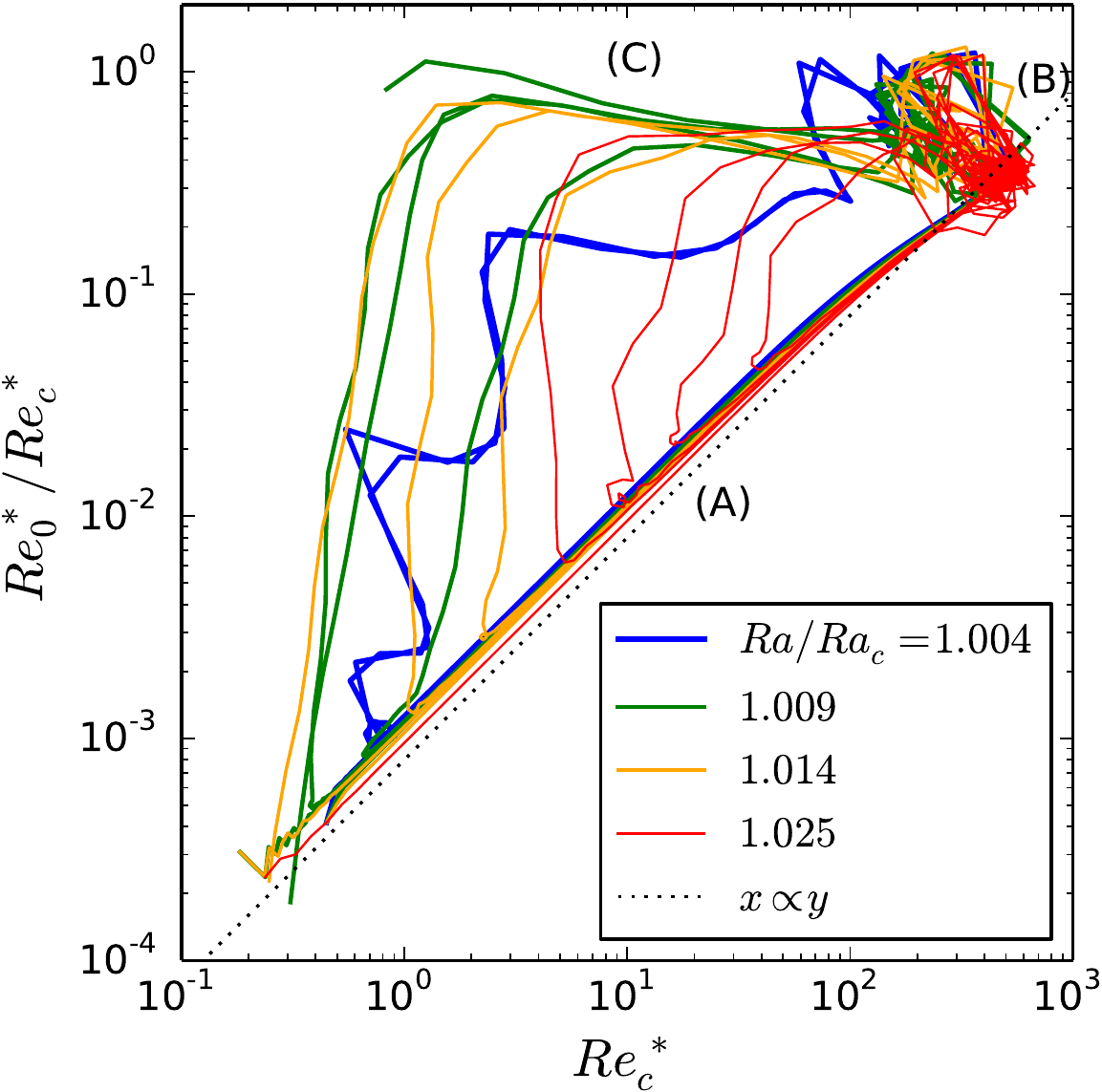}}
   \subfigure[]{\label{fig:phase_theta_int}
   \includegraphics[clip=true,width=7cm]{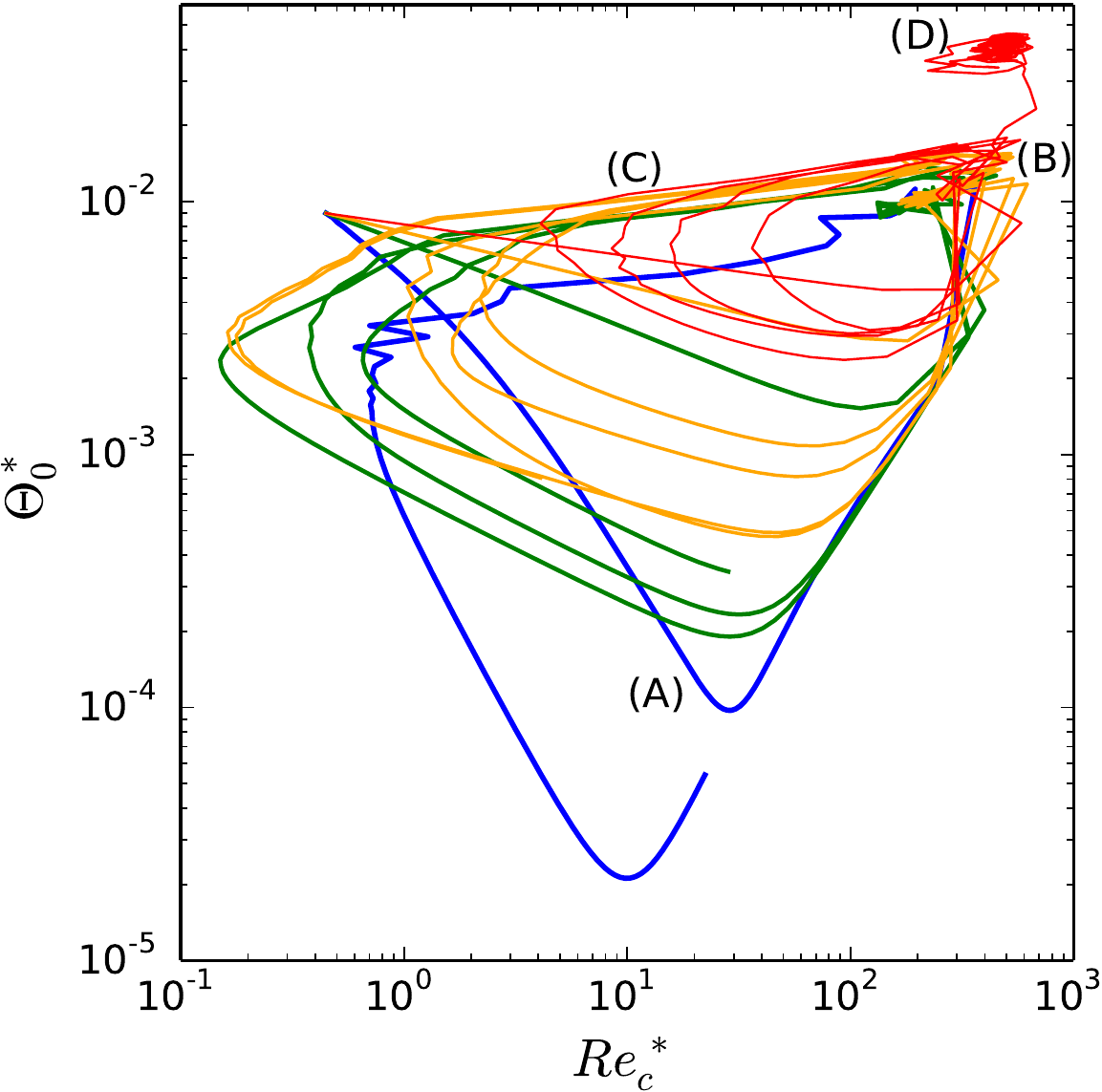}}
   \caption{Phase diagram of (a) $\uzrms^{\ast}/\ucrms^{\ast}$ versus $\ucrms^{\ast}$ 
   and (b) $\trms$ versus $\ucrms^{\ast}$ for the same time series as figure~\ref{fig:t_En_int}.}
\end{figure}

This behaviour is similar to nonlinear oscillations such as relaxation oscillations found in many dynamical systems \citep{Fau98}. 
These oscillations are quasi-periodic and may be better understood using phase diagrams.
Figure~\ref{fig:phase_int} shows the phase diagram of the ratio of the zonal to convective Reynolds numbers, 
$\uzrms^{\ast}/\ucrms^{\ast}$, versus the convective Reynolds number, $\ucrms^{\ast}$, for the same time series as figure~\ref{fig:t_En_int}. 
Figure~\ref{fig:phase_theta_int} shows the phase diagram for the rms axisymmetric temperature, $\trms$,
versus $\ucrms^{\ast}$.
During the exponential growing phase $\ucrms^{\ast}$  (labelled (A)),
all trajectories follow the same path and $\uzrms^{\ast}$ increases as the square of $\ucrms^{\ast}$, in agreement with weakly nonlinear 
analysis near the onset of convection (see \S\ref{sec:E5}). 
The system reaches a maximum (B), where $\uzrms^{\ast}$ has a similar amplitude to $\ucrms^{\ast}$. 
A dissipative phase (C) then follows during which
$\uzrms^{\ast}$ and $\ucrms^{\ast}$ initially decay at the same rate.
The three quantities,  $\ucrms^{\ast}$, $\uzrms^{\ast}$ and $\trms$, all start to decay at the same time. While $\ucrms^{\ast}$ and $\uzrms^{\ast}$
enter together the growing phase (A), the decay of $\trms$ persists for longer.
The three phases (A, B, C) are fairly similar for all the Rayleigh numbers, although the system can follow different trajectories
during the dissipative phase. 
For the largest Rayleigh number ($\Ra=1.025\Ra_c$), the system exits the oscillating regime 
to reach a stagnation point (D), which is clearly distinct from the phase (B) according to the measurements of $\trms$. 
During the phase (D), the characteristics of the flow are similar than the ones described in \S\ref{sec:E8} for $\Ek=10^{-8}$,
so this saturated regime corresponds to the strong branch of convection.

Relaxation oscillations are a well-known phenomenon in rotating spherical convection for Rayleigh
numbers a few times above onset \citep[\eg][]{Gro01,Mor04,Tee12}. 
However, unlike in these higher $\Ra$ relaxation oscillations, which are caused by the disruption of the convection
by the zonal flow and the axisymmetric temperature gradient \citep{Tee12}, we do not observe here a time lag between
the maximum of $\uzrms^{\ast}$ and $\ucrms^{\ast}$, so the oscillations of figure~\ref{fig:t_En_int} are produced by a different mechanism.

\begin{figure}
\centering   
   \includegraphics[clip=true,height=5cm]{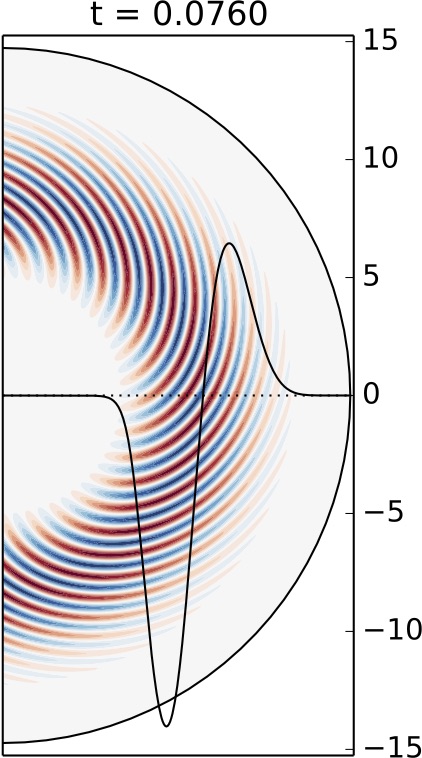}
   \includegraphics[clip=true,height=5cm]{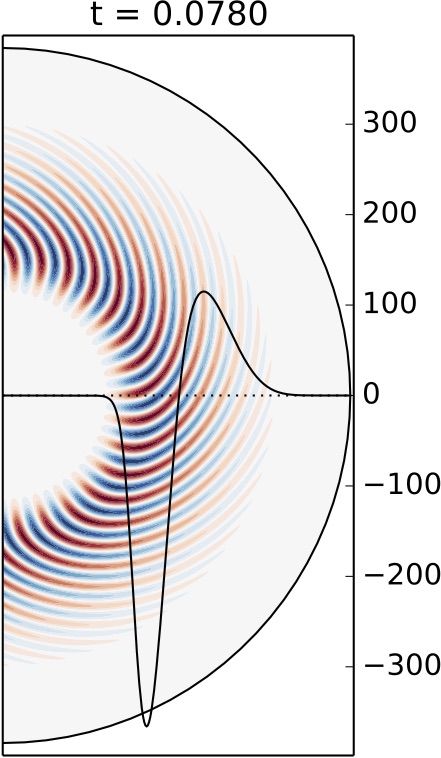}
   \includegraphics[clip=true,height=5cm]{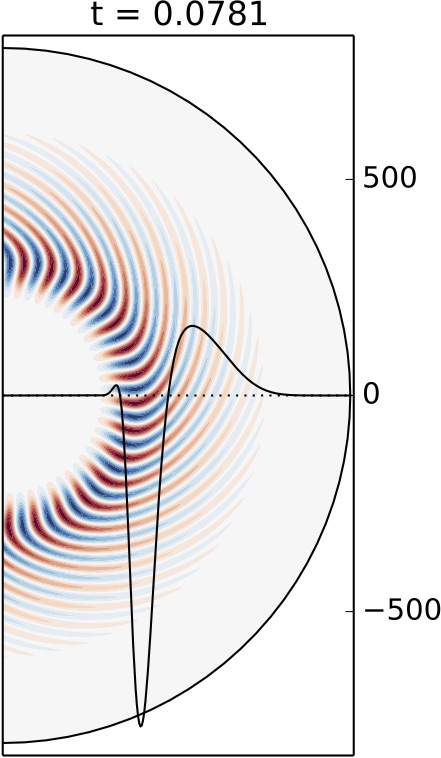}
   \includegraphics[clip=true,height=5cm]{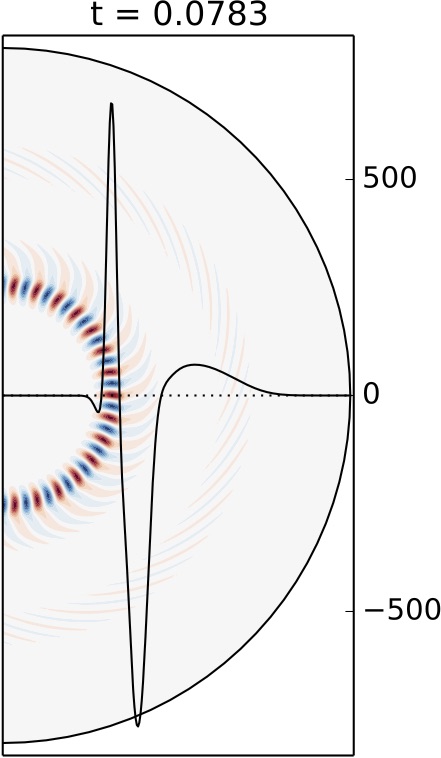}
   \includegraphics[clip=true,height=5cm]{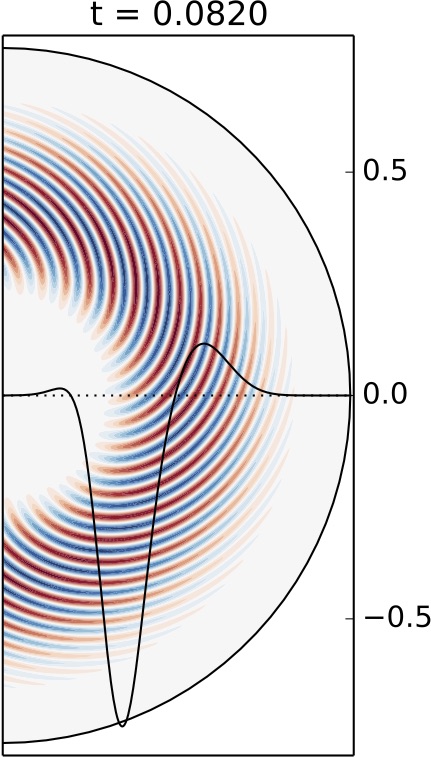}
   \caption{Snapshots of the streamfunction and the zonal flow for $E = 10^{-7}$, $Pr=0.1$ and 
   $Ra =1.004\Ra_c$. Between the first and last snapshots, a time corresponding to approximately $10^4$ rotation periods has elapsed.}
\label{fig:Psi_E7P1R490}
\end{figure}

Figure~\ref{fig:Psi_E7P1R490} shows a series of snapshots of the streamfunction (colour) and the zonal flow
(solid black line) during the different phases of the nonlinear oscillation for $\Ra=1.004\Ra_c$. 
During the growing phase (A) ($t=0.076$), the flow consists
of a thermal Rossby wave and a double zonal jet, both developing around the radius $s=0.5$, reminiscent of the solution on the weak branch. 
When the system reaches the maximum (B) ($t=0.078$), the thermal Rossby wave and the zonal flow drift inwards. 
A narrow prograde zonal jet starts to develop around $s=0.35$ ($t=0.0781$) and rapidly 
gains a very large amplitude ($t=0.0783$). The non-axisymmetric flow is maximum inside the narrow 
newly formed prograde jet and keeps the same azimuthal wavenumber as the thermal Rossby wave.  
The amplitude of both the zonal and non-axisymmetric flows
then decays rapidly (phase (C)). 
A new oscillation begins with the growth of a thermal Rossby wave and its associated double zonal jet
around the radius $s=0.5$, while the zonal flow of the past oscillation finishes to be dissipated ($t=0.082$).
The inward migration of the non-axisymmetric and zonal flows during the phase (B)
is probably the cause for the disruption of the convection and the oscillatory behaviour.
The mechanism responsible for the inward migration remains unclear as the competitive effects 
that either inhibit ($\beta$ effect) or promote (gravity along $s$) convection depend on radius.  

\begin{figure}
\centering
   \subfigure[]{
   \includegraphics[clip=true,height=6.5cm]{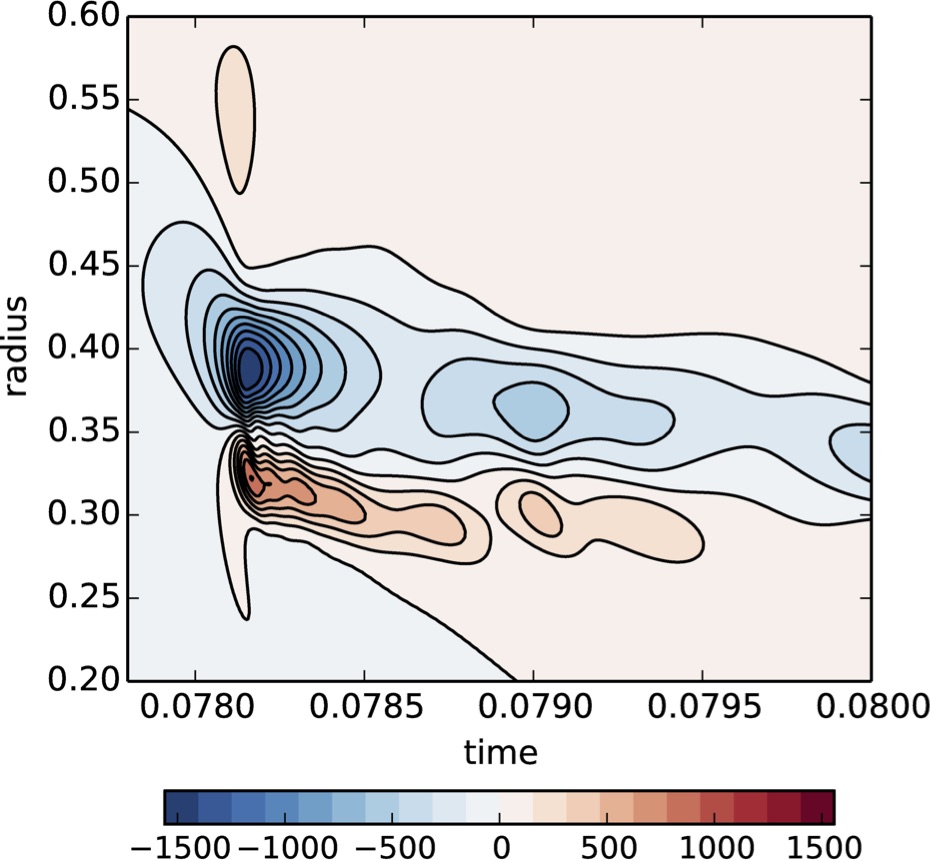}}
    \subfigure[]{
   \includegraphics[clip=true,height=6.5cm]{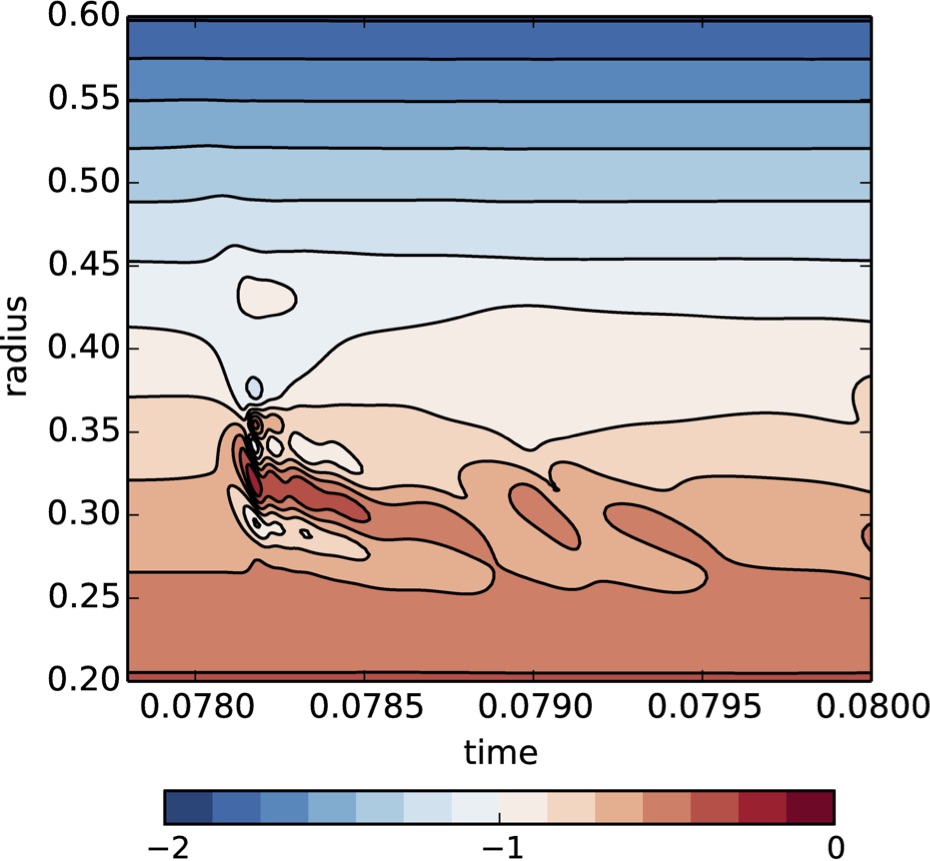}}
   \caption{Space-time diagram of (a) $\zw$ and (b) $\Delta$ for $E = 10^{-7}$, $Pr=0.1$ and $Ra =1.004\Ra_c$.}
\label{fig:Kuo}
\end{figure}

We note that a secondary effect might arise from the inward migration of the zonal flow in the form of a shear instability.
Indeed, the Rossby wave that is confined within the intense narrow prograde jet ($t=0.0783$ in figure~\ref{fig:Psi_E7P1R490})
could have been excited by such an instability. 
A sufficient and necessary condition for the barotropic instability of a shear flow in a inviscid Boussinesq
fluid is that the quantity \mbox{$\Delta=2\beta\Ek^{-1}-d\moyp{\vorz}/ds$}, where \mbox{$\moyp{\vorz}=d\zw/ds+\zw/s$}, 
changes sign at some radius \citep{Kuo49,Ing82}.
\cite{Gue12a} showed that the threshold of the instability obtained with numerical simulations at finite $\Ek$ approaches asymptotically
the inviscid theoretical prediction as the Ekman number is decreased and a good quantitative agreement is obtained at $\Ek=10^{-7}$. 
Figure~\ref{fig:Kuo} shows the space-time diagram of $\zw$ and $\Delta$ between the time $t=0.0778$ and $t=0.08$ 
for the same simulation as figure~\ref{fig:Psi_E7P1R490}. 
$\Delta$ is always negative, although it comes close to zero when the narrow prograde jet is at its maximum.
It is nonetheless possible that the linear stability criteria is violated, but this cannot be strictly identified in the nonlinear simulation
because the zonal flow would be rapidly modified by the growth of the shear instability. 
We emphasize that if the zonal flow is shear unstable, it is a secondary effect and not the cause of the inward migration.

We find similar nonlinear oscillations in the vicinity of the linear onset of convection for $\Ek=10^{-6}$ and $\Pran=10^{-1}$, 
but in a narrower window of Rayleigh numbers (between $\Ra=\Ra_c$ and $\Ra=1.0015\Ra_c$). 
For larger Rayleigh numbers, the convection is similar to the solution found on the weak branch 
with thermal Rossby waves. This therefore indicates that the nonlinear oscillations are not a consequence of the 
existence of a strong branch of convection.

\section{Discussion}
\label{sec:ccl}

We have studied rotating thermal convection driven by internal heating in a full sphere
near the onset of convection for values of the Prandtl number relevant for liquid metals ($\Pran\in[10^{-2},10^{-1}]$)
and low Ekman numbers ($\Ek\in[10^{-8},10^{-5}]$). We have used
an hybrid numerical model that couples a quasi-geostrophic approximation
for the velocity to a fully 3D temperature field. 
The model includes Ekman pumping to mimic no-slip boundary conditions.
Our main finding is the identification of two distinct branches of rapidly-rotating convection: 
(i) a weak branch for $\Ek\geq\mathit{O}(10^{-6})$
that is linked to the linear onset of convection (at $\Ra=\Ra_c$) by a supercritical bifurcation 
and (ii) a strong branch for lower Ekman numbers that is discontinuous at the onset and can only be reached by using finite amplitude perturbations as initial conditions.
On the weak branch for $\Ra$ just above $\Ra_c$, the flow consists of the interaction of thermal Rossby waves and a zonal flow
with a double jet structure, and has been extensively documented in the literature 
\citep[\eg][]{Bus82,Zha92}. 
The amplitude of the zonal flow 
on this branch near onset is smaller than the amplitude of the convective (\ie non-axisymmetric) velocity.
On the strong branch, the flow contains a wide range of length scales and its Reynolds number
$\urms$ exceeds $10^3$. The zonal flow has multiple jets for $\Ek\leq10^{-7}$ and its amplitude
is comparable to or larger than the convective velocity, even near the nonlinear onset of convection.
For $\Ek=10^{-5}$, the transition from quasi-steady convection to time-dependent convection 
that produces an abrupt kink in the evolution of $\urms$ and $\Nu$ as a function of $\Ra$ (see figures~\ref{fig:E5_urms}-\ref{fig:E5_Nu})
corresponds perhaps to the continuous transition from the weak to the strong branch. This transition has been observed
in previous 3D models \citep{Til97,Sim03}, 
so the existence of the two branches of convection might not be specific
to the quasi-geostrophic model. 
When the viscous effects become smaller, the transition between the two branches becomes discontinuous leading to the existence of two distinct 
branches, which allows us to observe
a hysteresis loop for $\Ek=10^{-6}$ and $\Pran=10^{-2}$, where the onset of the strong branch occurs at 
$\Ra_s=1.09\Ra_c$.

\begin{figure}
\centering   
\includegraphics[clip=true,width=14cm]{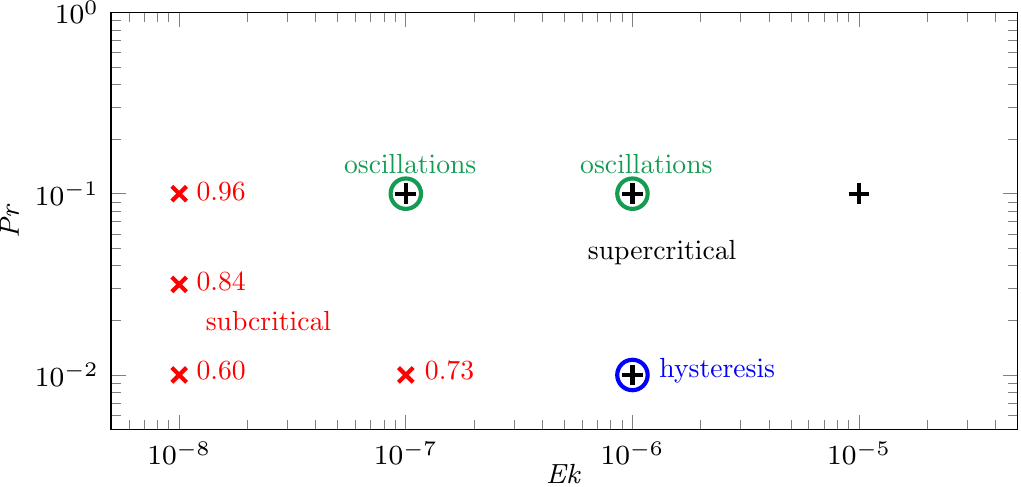}
   \caption{Regime diagram showing the location of the subcritical convection (red cross) 
   and supercritical bifurcation at the onset of nonlinear convection (black plus) identified
   with our hybrid model.  The value annotated next to the subcritical cases gives the ratio $\Ra_s/\Ra_c$.}
\label{fig:regime}
\end{figure}

For yet lower Ekman numbers, we find that the strong branch becomes 
subcritical with $\Ra_s < \Ra_c$. In this case, we did not find the weak branch that is continuous at the linear onset.
The parameters $(\Ek,\Pran)$ for which we observe subcritical convection are denoted by red crosses
in the regime diagram shown in 
figure~\ref{fig:regime}. The ratio $\Ra_s/\Ra_c$ is indicated next to each point and decreases
with the Ekman number and the Prandtl number. 
To the best of our knowledge, this is the first time that subcriticality is reported in thermal convection in a rotating sphere. 
It is possible that this behaviour is specific to the quasi-geostrophic model, 
as the nonlinear bifurcation might be sensitive to the treatment of the axial velocity for instance.
Nevertheless subcritical convection in the fully 3D system has been expected from the work of \citet{Sow77} in the asymptotic limit $\Ek\to0$
(see figure 3.2 of \citet{Pro94}).
\citeauthor{Sow77} found that weakly nonlinear solutions exist close to the critical Rayleigh number predicted by the local asymptotic theory, $\Ra_c^l$
\citep{Bus70}, which is smaller than the critical Rayleigh number of the global asymptotic theory, $\Ra_c^g$ \citep{Jon00}.
The discrepancy between local and global asymptotic theories increases at small Prandtl numbers (see Appendix \ref{sec:linear}),
so it is perhaps unsurprising that the subcriticality is amplified in this case in our simulations. 
For $\Pran=10^{-2}$, $\Ra_c^l/\Ra_c^g=0.03$, while $\Ra_c^l/\Ra_c^g=0.82$ for $\Pran=1$ \citep{Jon00}. 
For $\Ek=10^{-8}$ and $\Pran=10^{-2}$, we find that $\Ra_s/\Ra_c=0.60$, which could indicate that our simulations are still a long way from the asymptotic limit.

The region of intermediate Ekman numbers, between supercritical bifurcation at $\Ek\geq10^{-5}$ and subcritical
convection at $\Ek\lesssim10^{-7}$, displays interesting behaviours such as hysteresis.  
We also observe nonlinear oscillations for Rayleigh numbers just above the linear onset and up to $\Ra/\Ra_c=1.025$ for $\Ek=10^{-7}-10^{-6}$ and $\Pran=10^{-1}$.
For $\Ek=10^{-7}$, the dynamical regime is located on the strong branch for larger Rayleigh numbers, so in this case $\Ra_s\approx \Ra_c$.  
For $\Ek=10^{-6}$ however, the solution at larger $\Ra$ is similar to the convection at larger Ekman numbers.
In both cases, the nonlinear oscillations are due to an inward migration of the convective region, whose cause remains unexplained (see figure~\ref{fig:Psi_E7P1R490}).

The results presented in this paper rely on a quasi-geostrophic model of the velocity and a 3D model of the temperature, which allowed
us to compute small Ekman numbers currently out of reach of 3D models \citep[\eg][]{Miy10,Yad16, Mat16}. 
Assuming that our results are not specific
to the hybrid model, it would be difficult to observe subcriticality for 3D models in the near future, but it might be possible to observe 
the discontinuous transition between the two branches at low enough Prandtl numbers
for $\Ek=10^{-6}$ for instance.

In order to extrapolate the results of numerical models to natural bodies such as the Earth's core, the models must capture the proper
convection branch. Indeed here, we find that the scaling of the Nusselt number with the parameters are significantly different 
on the weak and strong branches (see figure~\ref{fig:Nu_hyst} for the hysteresis). 
In this paper, we do not derive scaling laws for the strong branch for all the studied parameters 
because it is unclear whether the points actually
follow a power law for the small values of $\Ra/\Ra_c$ studied here. 
The zonal flows and the scaling of the Reynolds and Nusselt numbers
at higher Rayleigh numbers will be addressed in a forthcoming study.

Many questions remain unanswered in our study and, in particular,
we have not identified the physical mechanisms leading to subcritical convection in our simulations
and controlling the value of $\Ra_s$.
The zonal flow and/or the axisymmetric temperature are probably essential for the
subcriticality \citep{Pla08}.
Whether the temperature needs to be treated in 3D to observe this behaviour is unclear.
The role played by the type of thermal heating also remains an open question.
Following the argument that the distinction between the local and global asymptotic theories yields
the existence of subcritical convection, \cite{Gil06} argue that the case of differential heating might well always
be supercritical because no such distinction exists in this case.

The context of the present work is the study of the fluid dynamics in planetary liquid cores. 
It is of interest to note that, in this context, the possibility of the subcritical behaviour of rapidly-rotating convection 
due to the action of a magnetic field is an active area of research \citep[\eg][]{Sre11,Dor16}. 
In magnetoconvection, where a magnetic field is imposed, the Lorentz force can counteract the inhibiting action of the Coriolis force
on the convection and leads to a reduced onset of convection \citep{Cha61,Elt72, Fea79}. 
In dynamo simulations, the magnetic field is sustained by the convective flow and saturates due to the feedback of Lorentz forces
on the flow. When a strong magnetic field is sustained, the Lorentz forces can strongly affect the flow and it is then possible 
that the Rayleigh number required to maintain the dynamo falls below the Rayleigh number required to excite the dynamo in the first place at the
dynamo onset (which occurs at $\Ra>\Ra_c$). Examples of this subcritical behaviour below the dynamo onset
have been reported by \citet{kua08,Mor09,Sre11,Hor13}, and have been used as a possible explanation for the sudden termination of the martian dynamo. 
However in these examples of convectively-driven dynamos in spherical geometry, the Rayleigh number always exceeds $\Ra_c$. 
As far as we know, the only example of a dynamo simulation operating below $\Ra_c$ was found by \citet{Ste04} in a planar model
of rotating convection.
It would be of great interest to study the dynamo action produced by the flows on the strong hydrodynamical branch identified in this
paper, and whether the Lorentz force can push the system to remain convective at yet smaller Rayleigh numbers as suggested for instance 
by \citet{Rob88}.

\section*{Acknowledgements}
CG was supported by the Natural Environment Research Council under grant NE/M017893/1.
PC acknowledges the Agence Nationale de la Recherche for supporting this project under grant ANR TuDy.
This work was undertaken on ARC1 and ARC2 of the HPC facilities at the University of Leeds,
on TOPSY of the HPC facilities at Newcastle University and on the Froggy platform of the CIMENT infrastructure (\url{https://ciment.ujf-grenoble.fr}).
The CIMENT infrastructure is supported by the Rh\^one-Alpes region (GRANT CPER07\_13 CIRA), 
the OSUG@2020 labex (reference ANR10 LABX56) and the Equip@Meso project (reference ANR-10-EQPX-29-01).
ISTerre is part of Labex OSUG@2020 (ANR10 LABX56).
We are grateful to Emmanuel Dormy, Andrew Soward, Toby Wood and the Geodynamo group in Grenoble
for helpful discussions and to the anonymous referees for suggestions that have improved the manuscript.

\appendix
\section{Linear onset of convection}
\label{sec:linear}

The onset of thermal convection in a rotating sphere with internal heating 
in the asymptotic limit $\Ek\ll1$ and $\Ek/\Pran\ll1$ was studied analytically by \citet{Rob68,Bus70,Yan92,Jon00}.
In this limit, the most unstable mode of convection develops in the form of columnar structures aligned
with the rotation axis, which are called thermal Rossby waves. They develop around a radius $s\approx 0.5$. 
Using a local perturbative study, \citet{Bus70} determined the analytical dependence of the critical parameters at the onset 
of convection as function of $\Ek$ and $\Pran$,
\begin{eqnarray}
 \Ra_c^l&=& C_r \pl \frac{\Pran}{1+\Pran}\pr^{4/3} \Ek^{-4/3} ,
\label{eq:B_Rac} \\
 m_c^l&=& C_m \pl \frac{\Pran}{1+\Pran} \pr^{1/3} \Ek^{-1/3} ,
\label{eq:B_mc}\\
 \omega_c^l &=& C_{\omega} \pl \Pran(1+\Pran)^2\pr^{-1/3}  \Ek^{-2/3} ,
\label{eq:B_wc}
\end{eqnarray}
where  $\Ra_c^l$ is the critical Rayleigh number, $m_c^l$ the marginally stable azimuthal wavenumber, 
$\omega_c^l$ its frequency, and $C_r$, $C_m$ and $C_{\omega}$ are constant.
However, \citet{Sow77} showed that small disturbances decay with time for values of the Rayleigh number just above 
the local critical Rayleigh number (\ref{eq:B_Rac}). 
 \citet{Jon00} calculated the true value of the critical Rayleigh number from the global asymptotic theory, $\Ra_c^g$, and found that it is
 significantly larger than $\Ra_c^l$.
 
In order to benchmark our hybrid model, we have performed calculations with a linearised version of the code
and compared the results with the values of the  global asymptotic theory of \citet{Jon00}
for $\Pran\in[10^{-2},1]$. 
We study the onset of the thermal Rossby waves, so we restrict our study to the domain 
of small Ekman numbers where they are 
preferred over the equatorially-attached modes,
\ie $\Ek\leq10^{-4}$ for $\Pran=10^{-1}$ and  $\Ek\leq10^{-6}$ for $\Pran=10^{-2}$ (see \S\ref{sec:intro}).
We also compare the dependence 
of the critical parameters on $\Pran$ with the power laws of the local theory (\ref{eq:B_Rac})-(\ref{eq:B_wc}).
We calculated the linear onset for $\Pran=1$ at $\Ek\in[10^{-6},10^{-5}]$
to compare our results with published numerical results obtained with a fully 3D model \citep{Dor04} 
and a quasi-geostrophic description, where the temperature is also treated in 2D (hereafter denoted QG-2D) \citep{Aub01b}.
In these previous studies, the critical parameters for convection driven by internal heating were only 
published for $\Pran=1$ and the model includes an inner core of radius $0.35$.
The boundary conditions for the 3D model of \citet{Dor04} are no-slip,
while the QG-2D model of  \citet{Aub01b} neglects the Ekman pumping.

\begin{table}
  \begin{center}
\def~{\hphantom{0}}
  \begin{tabular}{lccccc}
      $\Ek$  & $\Pran$   &   $\Ra_c$ & $m_c$ & $\omega_c$ & $\Ra_c/\Ra_c^g$ \\[3pt]
      \hline
      $10^{-5}$  & $1$ & $4.039\times 10^7$ & $14$ & $1.44 \times 10^3$ & $1.68$\\
      $10^{-6}$ & $1$ & $8.374\times10^8$ & $32$ & $7.36 \times 10^3$ & $1.61$\\
      \hline
      $10^{-5}$ & $10^{-1}$ & $1.416\times 10^7$ & $8$ & $3.85\times10^3$ & $2.14$\\
      $10^{-6}$ & $10^{-1}$ & $2.736 \times 10^8$ & $18$ & $1.88\times10^4$ &$1.92$\\
      $10^{-7}$ & $10^{-1}$ & $5.032\times10^9$ & $44$ & $9.08\times10^4$ & $1.64$\\
      $10^{-8}$ & $10^{-1}$ & $7.759\times10^{10}$ & $109$ & $4.73\times10^5$ & $1.17$\\
      \hline
      $10^{-6}$ & $10^{-2}$ & $9.389 \times 10^7$ & $9$ & $3.95\times10^4$ & $1.56$\\
      $10^{-7}$ & $10^{-2}$ & $1.818 \times 10^9$ & $20$ & $1.90\times10^5$ & $1.41$\\ 
      $10^{-8}$ & $10^{-2}$ & $2.962\times10^{10}$ & $53$ & $8.08\times10^5$ & $1.06$\\
  \end{tabular}
  \caption{Critical parameters at the linear onset of convection calculated with a linearised version of the hybrid QG-3D numerical code.}
  \label{tab:critpar}
  \end{center}
\end{table}

\begin{figure}
\centering   
   \includegraphics[clip=true,width=\textwidth]{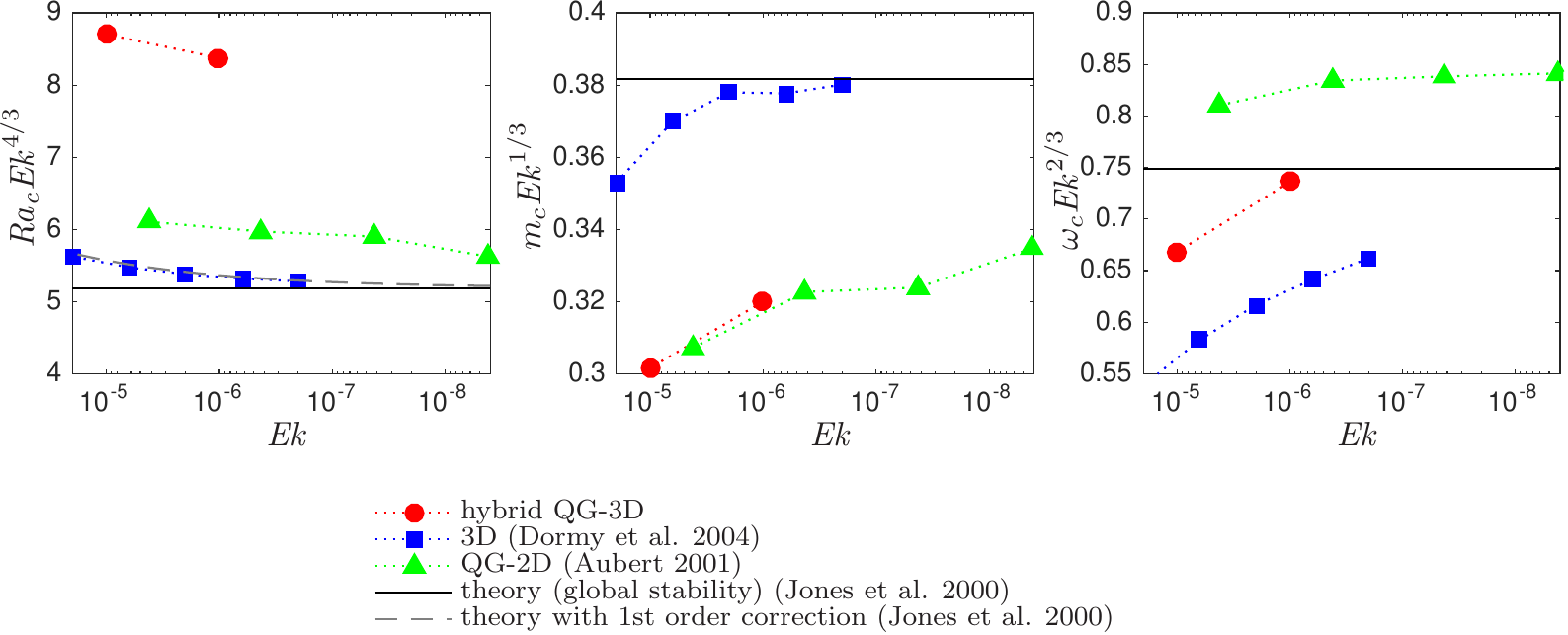}
   \caption{Critical parameters for the linear onset of convection as a function of $\Ek$ for $\Pran=1$
   computed with our model (hybrid QG-3D), with
   the 3D model of \citet{Dor04} and the quasi-geostrophic model of \citet{Aub01b}. These numerical results are compared
   with the asymptotic results of \citet{Jon00}. The dashed gray line represents the asymptotic
   results including first-order correction \citep{Jon00}.}
\label{fig:onset_vs_Ek}
\end{figure}

The critical parameters calculated with our code $(Ra_c, m_c, \omega_c)$ are given in table~\ref{tab:critpar}.
As the Ekman number is decreased, $\Ra_c$ gets closer to the asymptotic value $\Ra_c^g$ for all $\Pran$.
The values of the critical parameters are plotted as a function of $\Ek$ in figure~\ref{fig:onset_vs_Ek} for $\Pran=1$.
The three critical parameters computed with our hybrid model tend towards the asymptotic values of
the global stability analysis when $\Ek$ decreases.
The critical Rayleigh number of our model is higher than the values obtained with both the 3D and QG-2D models. 
This is expected because of the stabilising effect of the axial rigidity imposed by the quasi-geostrophic formulation, while the 3D solution keeps a large-scale $z$-dependence \citep{Gil06}.
Furthermore, our hybrid approach includes the axial diffusion of the temperature,
which is neglected in the QG-2D model. The $z$-average of the buoyancy force is thus weaker in our hybrid
approach, so $\Ra_c$ is necessarily larger.   
The marginally stable mode $m_c$ is smaller in our hybrid model compared with the asymptotic results. 
Again, this is consistent with the axial rigidity of the quasi-geostrophic flow, which 
requires the convection to first develop at a slightly smaller critical radius, thereby selecting a smaller $m_c$. 

\begin{figure}
\centering   
   \includegraphics[clip=true,width=\textwidth]{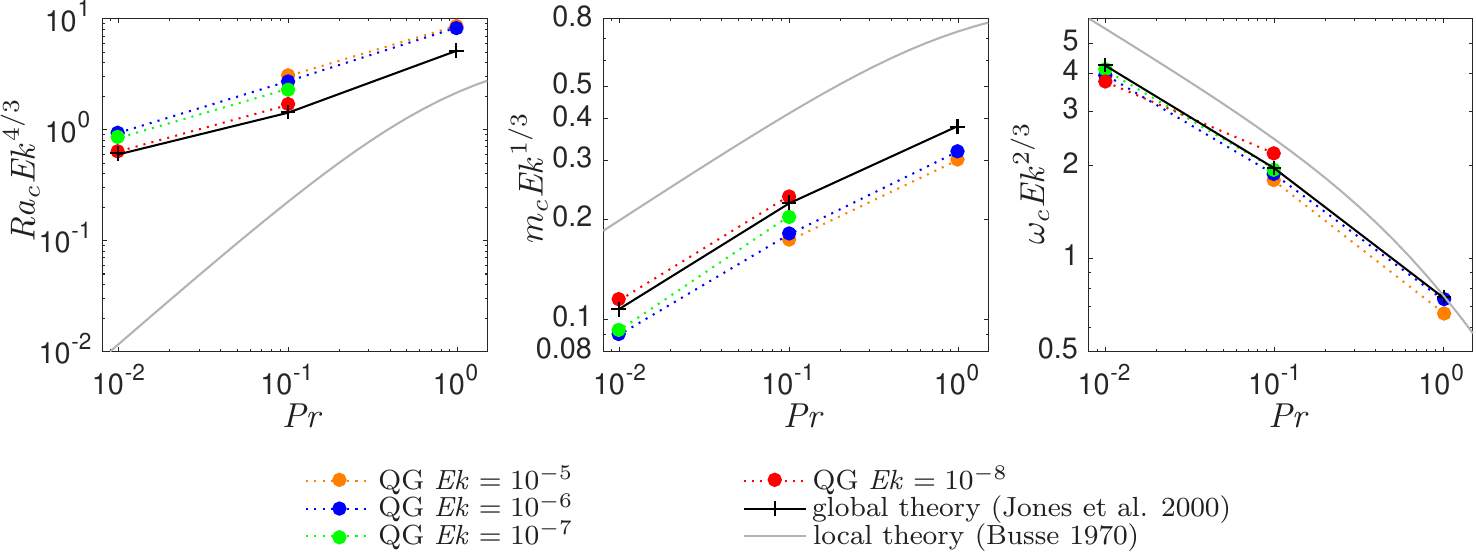}
   \caption{Critical parameters for the linear onset of convection as a function of $\Pran$ computed with our 
   hybrid QG model and compared with the asymptotic results of \citet{Jon00} (global theory) 
   and \citet{Bus70} (local theory).}
\label{fig:onset_vs_Pr}
\end{figure}

The critical parameters $(Ra_c, m_c, \omega_c)$ are plotted as a function of $\Pran$ in figure~\ref{fig:onset_vs_Pr}.
Our hybrid model follows well the trend of the global asymptotic theory for the three critical parameters.
Convection at smaller $\Pran$ develops on larger scales at the onset. 
Accordingly, the critical frequency is larger \citep[\eg][]{Jon07}.
These results agree reasonably well with the trends given by
the power laws of the local theory for $m_c^l$ and $\omega_c^l$ ((\ref{eq:B_mc})-(\ref{eq:B_wc})) when $\Pran<0.1$, but
the prefactors differ significantly.
The critical Rayleigh number given by both our numerical results and the global asymptotic results
differs greatly from the prediction of the local theory ($\Ra_c^l \propto \Pran^{4/3}$ 
when $\Pran\ll1$), as already observed by \citet{Zha92}. 
The discrepancy with $\Ra_c^l$ (\ref{eq:B_Rac}) increases as the Prandtl number
decreases: the difference between local and global critical Rayleigh number is nearly two orders of magnitude for $\Pran=10^{-2}$.
\citet{Zha92} attributed this discrepancy to the tilt of the thermal Rossby waves, which 
is more pronounced for small $\Pran$.

\end{document}